%% file: pldi18.tex
\documentclass[sigplan,screen]{acmart}

\usepackage{booktabs}
\usepackage{subcaption}

\usepackage{amsmath}
\usepackage{amsthm}
\usepackage{amssymb}
\usepackage{array}
\usepackage{flushend}
\usepackage{graphicx}
\usepackage{listings}
\usepackage{multirow}
\usepackage{placeins}
\usepackage{stmaryrd}
\usepackage[normalem]{ulem}
\usepackage{url}
\usepackage{xcolor}
\usepackage{xfrac}
\usepackage{xkeyval}
\usepackage{xspace}

\input{macros}

\setcopyright{rightsretained}
\acmPrice{}
\acmDOI{10.1145/3192366.3192387}
\acmYear{2018}
\copyrightyear{2018}
\acmISBN{978-1-4503-5698-5/18/06}
\acmConference[PLDI'18]{39th ACM SIGPLAN Conference on Programming Language Design and Implementation}{June 18--22, 2018}{Philadelphia, PA, USA}

\begin{document}

\title[Automated Clustering and Program Repair for Introductory \dots]{Automated Clustering and Program Repair for Introductory Programming Assignments}
\titlenote{Supported by the Austrian National Research Network S11403-N23 (RiSE) of the Austrian Science Fund (FWF).}

\author{Sumit Gulwani}
\affiliation{
  \institution{Microsoft Corporation}
  \country{USA}
}
\email{sumitg@microsoft.com}

\author{Ivan Radi\v{c}ek}
\affiliation{
  \institution{TU Wien}
  \country{Austria}
}
\email{radicek@forsyte.at}

\author{Florian Zuleger}
\affiliation{
  \institution{TU Wien}
  \country{Austria}
}
\email{zuleger@forsyte.at}



\input{abstract}

\maketitle

\input{intro}

\input{overview}

\input{model}

\input{matching}

\input{repair}

\input{experiments}

\input{related}

\input{futurework}

\input{conclusion}

\appendix

\input{appendix-problems}

\input{appendix-code-examples}

\bibliography{pldi18}

\end{document}

%% file: macros.tex
\newcommand\keyw[1]{\texttt{#1}} 

\newcommand\clara{\textsc{Clara}\xspace}

\newcommand\clang{\textsc{C}\xspace} 
\newcommand\pylang{\textsc{Python}\xspace} 
\newcommand\dom{\mathit{dom}} 
\newcommand\inlinecode{\texttt} 
\newcommand\mvar{\mathit} 

\newcommand\parto{\rightharpoonup}
\newcommand\pts{\mapsto}

\newcommand\EX[1]{\textbf{#1}}

\newcommand\ops{O} 
\newcommand\consts{K} 
\newcommand\var{v} 
\newcommand\vars{V} 
\newcommand\expr{e} 
\newcommand\exprs{E} 
\newcommand\prog{P} 
\newcommand\progb{Q} 

\newcommand\loc{\ell} 
\newcommand\locs{L} 
\newcommand\init{\mvar{init}}
\newcommand\linit{\loc_{\init}}
\newcommand\true{\keyw{true}}
\newcommand\false{\keyw{false}}

\newcommand\blabel{\mathcal{U}}
\newcommand\succfun{\mathcal{S}}
\newcommand\cond{\keyw{?}} 
\newcommand\compval{d} 
\newcommand\compdom{D} 
\newcommand\undeff{\bot} 
\newcommand\mem{\sigma} 
\newcommand\mems{\Sigma} 
\newcommand\inmem{\rho} 
\newcommand\semfun[1]{\!\left \llbracket #1 \right \rrbracket\!} 

\newcommand\retvar{\mathit{return}}

\newcommand\trace{\gamma} 

\newcommand\lend{\mvar{end}}
\newcommand\specvars[1]{{#1}^{\sharp}}

\newcommand\isom{\pi}

\newcommand\progsim{\sim}
\newcommand\simrel{\tau}
\newcommand\specs{\mathcal{\prog}}
\newcommand\classes{\mathcal{C}}

\newcommand\ins{I}

\newcommand\cexprs{\mathcal{E}}
\newcommand\esim{\simeq}
\newcommand\traces{\Gamma}

\newcommand\lpre{\loc_{\mathit{before}}}
\newcommand\lcond{\loc_{\mathit{cond}}}
\newcommand\lloop{\loc_{\mathit{loop}}}
\newcommand\lpost{\loc_{\mathit{after}}}

\newcommand\constrs{\mathcal{C}}
\newcommand\objective{\mathcal{O}}

\newcommand\exprvars{\mathcal{V}}
\newcommand\ints{\mathcal{I}}
\newcommand\assgn{\mathcal{A}}
\newcommand\repair{r}

\newcommand\parmap{\omega}

\newcommand\impl{{\mvar{impl}}}

\newcommand\pimpl{{\prog_\impl}}

\newcommand\nomod{\bullet}
\newcommand\repaired{\mathit{repaired}}
\newcommand\PM{\mathit{LR}}

\newcommand\ignore[1]{}

\newcommand\figlabel[1]{\label{fig:#1}}
\newcommand\figref[1]{Fig.~\ref{fig:#1}}
\newcommand\seclabel[1]{\label{sec:#1}}
\newcommand\secref[1]{\S\ref{sec:#1}}
\newcommand\tablabel[1]{\label{tab:#1}}
\newcommand\tabref[1]{Table~\ref{tab:#1}}
\newcommand\deflabel[1]{\label{def:#1}}
\newcommand\defref[1]{Def.~\ref{def:#1}}

\newcommand\lemlabel[1]{\label{lem:#1}}
\newcommand\lemref[1]{Lemma~\ref{lem:#1}}

\newenvironment{myitemize}{\begin{list}{$\bullet$}
{\setlength{\topsep}{1mm}\setlength{\itemsep}{0.25mm}
\setlength{\parsep}{0.1mm}
\setlength{\itemindent}{0mm}\setlength{\partopsep}{0mm}
\setlength{\labelwidth}{15mm}
\setlength{\leftmargin}{4mm}}}{\end{list}}

\definecolor{mygreen}{rgb}{0,0.6,0}
\definecolor{mygray}{rgb}{0.5,0.5,0.5}
\definecolor{mymauve}{rgb}{0.58,0,0.82}
\newcommand\includecode[2]{{{\minipage{#2\textwidth}\lstinputlisting{#1}\endminipage}}}



%% file: abstract.tex
\begin{abstract}
Providing feedback on programming assignments is a tedious task for the
instructor, and even impossible in large Massive Open Online Courses with
thousands of students.
Previous research has suggested that program repair techniques can be used to
generate feedback in programming education.
In this paper, we present a novel fully automated program repair algorithm for
introductory programming assignments.
The key idea of the technique, which enables automation and scalability, is to
use the existing correct student solutions to repair the incorrect attempts.
We evaluate the approach in two experiments:
(I)~We evaluate the number, size and quality of the generated repairs on 4,293
incorrect student attempts from an existing MOOC.
We find that our approach can repair 97\% of student attempts,
while 81\% of those are small repairs of good quality.
(II)~We conduct a preliminary user study on performance and repair usefulness
in an interactive teaching setting.
We obtain promising initial results (the average usefulness grade 3.4 on a
scale from 1 to 5), and conclude that our approach can be used in an
interactive setting.
\end{abstract}

%% file: intro.tex
\begin{figure}[h!]
  \centering
  \includegraphics[width=0.45\textwidth]{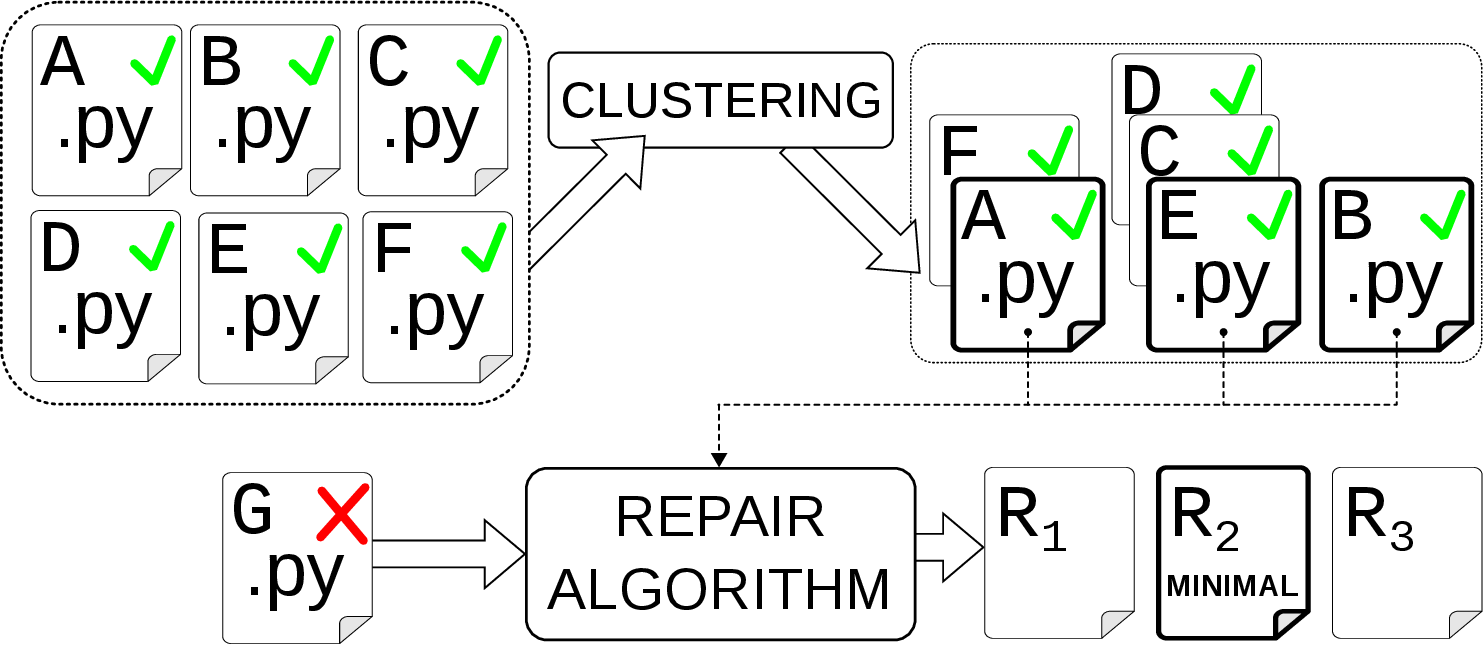}
  \caption{High-level overview of our approach.}
  \figlabel{Overview}
\end{figure}

\section{Introduction}\seclabel{intro}

Providing feedback on programming assignments is an integral part of a class on
introductory programming and requires substantial effort by the teaching
personnel.
This problem has become even more pressing with the increasing demand for
programming education, which universities are unable to meet (it is predicted
that in the US by 2020 there will be one million more programming jobs than
students~\cite{geekwire}).
This has given rise to several Massive Open Online Courses (MOOCs) that teach
introductory programming~\cite{masters:MOOCs}; the biggest challenge in such a
setting is scaling personalized feedback to a large number of students.

The most common approach to feedback generation is to present the student with
a failing test case; either generated automatically using test input generation
tools~\cite{tillmann13:teaching} or selected from a comprehensive collection of
representative test inputs provided by the instructor.  This is useful
feedback, especially since it mimics the setting of how programmers debug their
code.
However, this is not sufficient, especially for students in an introductory
programming class, who are looking for more guided feedback to make progress
towards a correct solution.

A more guided feedback can be generated from modifications that make a
student's program correct, using a program repair technique as pioneered by the
AutoGrader tool~\cite{singh:pldi13}.
Generating feedback from program repair is an active area of research: One line
of work focuses on improving the \emph{technical capabilites of program repair}
in introductory education~\cite{Refazer,Rivers17,qlose}, while another line of
research focuses on \emph{pedagogical questions} such as how to best provide
repair-based feedback to the students~\cite{LS17,HintsDesignSpace}.
In this paper we propose a new completely automated approach for repairing
introductory programming assignments, while sidelining the pedagogical
questions for future work.

\paragraph{Our approach}
The key idea of our approach is to use the \emph{wisdom of the crowd}: we use
the \emph{existing correct} student solutions to repair the \emph{new incorrect}
student attempts\footnote{We distinguish correct and incorrect attempts by
  running them on a set of inputs, and comparing their output to the expected
  output.  This is the standard way of assessing correctness of student
  attempts in most introductory programming classes.}.
We exploit the fact that MOOC courses already have tens of thousands of existing
student attempts; this was already noticed by~\citet{Drummond:2014}.

\figref{Overview} gives a high-level overview of our approach:
(A)~For a given programming assignment, we automatically \emph{cluster the
  correct student solutions} (\EX{A}-\EX{F} in the figure), based on a
notion of \emph{dynamic equivalence} (see \secref{CorrectClassification} for an
overview and \secref{clustering} for details).
(B)~Given an \emph{incorrect student attempt} (\EX{G} in the figure) we run the
\emph{repair algorithm} against all clusters, and then select a \emph{minimal
  repair} ($\mathbf{R_2}$ in the figure) from the generated repair candidates
($\mathbf{R_1}$-$\mathbf{R_3}$ in the figure).
The repair algorithm uses expressions from \emph{multiple correct solutions} to
generate a repair (see \secref{IncorrectRepair} for an overview and
\secref{repair} for details).

Intuitively, our clustering algorithm groups together similar correct
solutions.
Our repair algorithm can be seen as a generalization of the clustering approach
of correct solutions to incorrect attempts.
The key motivation behind this approach is as follows: to help the student,
with an incorrect attempt, our approach finds the set of most similar correct
solutions, written by other students, and generates the smallest modifications that
get the student to a correct solution.

\input{motiv-examples}

We have implemented the proposed approach in a tool called \clara and evaluated
it in two experiments:

(I) On 12,973 correct and 4,293 incorrect (total 17,266) student attempts from
an \emph{MITx} MOOC, written in \pylang, we evaluate the \emph{number},
\emph{size} and \emph{quality} of the generated repairs.
\clara is able to repair 97\% of student attempts, in 3.2s on average;
we study the quality of the generated repairs by manual inspection and find
that \emph{81\%} of the generated repairs are \emph{of good-quality} and the
size of the generated repair \emph{matches the size of the required changes} to
the student's program.
Additionally, we compare AutoGrader and \clara, on the same MOOC data.

(II) We performed a preliminary user study about the \emph{performance} and
\emph{usefulness} of \clara's repairs in an interactive teaching setting.
The study consisted of 52 participants who were asked to solve 6
programming assignments in \clang.
The participants judged the usefulness of the generated repair-based feedback
by 3.4 in average on a scale from 1 to 5.

\emph{Our experimental results demonstrate that \clara can, completely
  automatically, generate repairs of high quality, for a large number of
  incorrect student attempts in an interactive teaching setting for introductory programming problems.}

This paper makes the following contributions:
\begin{itemize}
\item We propose an algorithm to automatically cluster correct student
  solutions based on a dynamic program analysis.
\item We propose a completely automated algorithm for program repair of
  incorrect student attempts that leverages the \emph{wisdom of the crowd} in a
  novel way.
\item We evaluate our approach on a large MOOC dataset and show that
  \clara can repair almost all the programs while generating repairs of high quality.
\item We find in a real-time user study that \clara is sufficiently fast to be
  used in an interactive teaching setting and obtain promising preliminary
  results from the participants of the user study on the usefulness of the
  generated feedback.
\end{itemize}

\emph{Differences to our earlier Technical Report.}
Our approach first appeared as a technical report\footnote{\url{https://arxiv.org/abs/1603.03165v1}},
together with a publicly available implementation\footnote{\url{https://github.com/iradicek/clara}}.
While our core ideas are the same as in the technical report, we since have made several
improvements of which we state here the two most important:
(1)~The repair algorithm uses expressions from different correct solutions in a
cluster, as opposed to using only a single correct solution from a cluster.
(2)~We conducted a user study and added an extensive manual investigation of the generated repairs to the experimental evaluation.

\emph{Structure of the paper.}
In \secref{overview} we present an overview of our approach,
in \secref{model} we describe a simple imperative language,
used for formalizing the notions of matching and clustering in
\secref{clustering} and the repair procedure in \secref{repair}.
We discuss our implementation and the experimental evaluation in
\secref{experiments}, overview the related work in \secref{related},
discuss limitations and directions for future work in \secref{futurework},
and conclude in \secref{conclusion}.

%% file: motiv-examples.tex
\begin{figure*}[t]
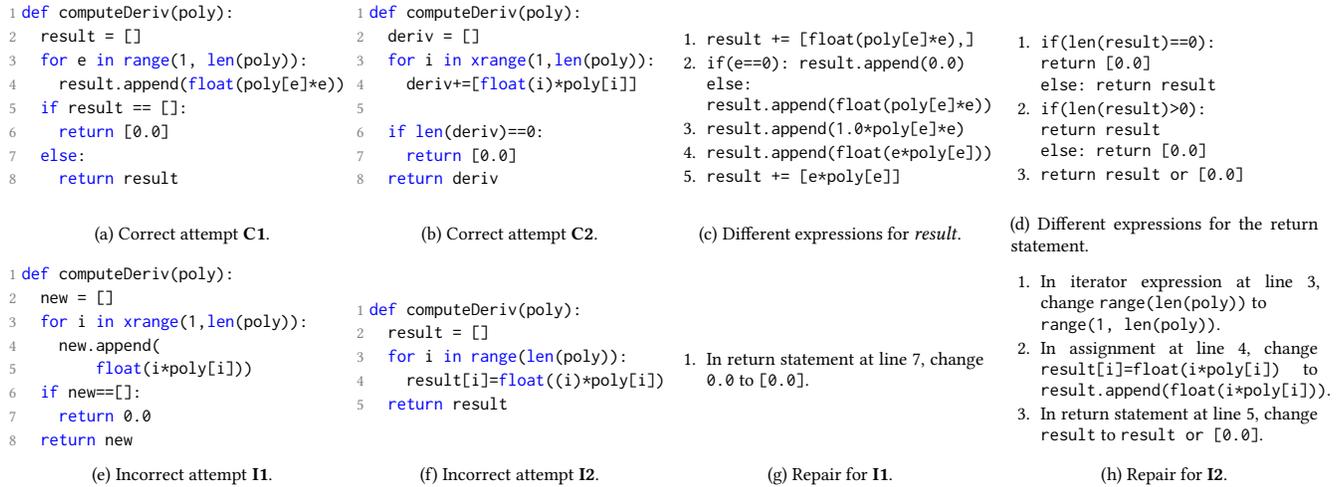

  \scriptsize
  \begin{tabular}{cccc}
    \includecode{examples/deriv-c1.py}{0.24} &
    \includecode{examples/deriv-c2.py}{0.21} &
    \begin{minipage}{0.23\textwidth}
      \vspace{0.3cm}
      \begin{myitemize}
      \item[1.] \inlinecode{result += [float(poly[e]*e),]}
      \item[2.] \inlinecode{if(e==0): result.append(0.0)} \\
        \inlinecode{else:} \\
        \inlinecode{result.append(float(poly[e]*e))}
      \item[3.] \inlinecode{result.append(1.0*poly[e]*e)}
      \item[4.] \inlinecode{result.append(float(e*poly[e]))}
      \item[5.] \inlinecode{result += [e*poly[e]]}
      \end{myitemize}
    \end{minipage}\vspace{0.2cm} &
    \begin{minipage}{0.23\textwidth}
      \vspace{0.3cm}
      \begin{myitemize}
      \item[1.] \inlinecode{if(len(result)==0):} \\
        \inlinecode{return [0.0]} \\
        \inlinecode{else: return result}
      \item[2.] \inlinecode{if(len(result)>0):}\\
        \inlinecode{return result} \\
        \inlinecode{else: return [0.0]}
      \item[3.] \inlinecode{return result or [0.0]}
      \end{myitemize}
    \end{minipage} \\
    (a) Correct attempt \textbf{C1}. &
    (b) Correct attempt \textbf{C2}. &
    (c) Different expressions for $\mathit{result}$. &
    \begin{minipage}{0.23\textwidth}
      (d) Different expressions for the return statement.
    \end{minipage}\\
    \includecode{examples/deriv-i1.py}{0.24} &
    \includecode{examples/deriv-i2.py}{0.21} &
    \begin{minipage}{0.23\textwidth}
      \vspace{0.3cm}
      \begin{myitemize}
        \item[1.] In return statement at line 7, change \inlinecode{0.0} to \inlinecode{[0.0]}.
      \end{myitemize}
    \end{minipage} &
    \begin{minipage}{0.23\textwidth}
      \vspace{0.3cm}
      \begin{myitemize}
      \item[1.] In iterator expression at line 3, change \inlinecode{range(len(poly))} to \\
        \inlinecode{range(1, len(poly))}.
      \item[2.] In assignment at line 4, change \inlinecode{result[i]=float(i*poly[i])} to \inlinecode{result.append(float(i*poly[i]))}.
      \item[3.] In return statement at line 5, change \inlinecode{result} to \inlinecode{result or [0.0]}.
      \end{myitemize}\vspace{0.2cm}
    \end{minipage} \\
    (e) Incorrect attempt \textbf{I1}. &
    (f) Incorrect attempt \textbf{I2}. &
    (g) Repair for \textbf{I1}. &
    (h) Repair for \textbf{I2}.
  \end{tabular}
  \caption{Motivation examples of real student attempts on the programming assignment \inlinecode{derivatives}.}
  \figlabel{MotivExamples}
\end{figure*}

%% file: overview.tex
\section{Overview}\seclabel{overview}

We discuss the high-level ideas of our approach on the student attempts to the
assignment \inlinecode{derivatives}:
``\emph{Compute and return the derivative of a polynomial function (represented
  as a list of floating point coefficients).  If the derivative is
  \inlinecode{0}, return \inlinecode{[0.0]}.}''
\figref{MotivExamples}~(a),~(b),~(e)~and~(f) show four student attempts to the
above programming assignment:
\EX{C1} and \EX{C2} are functionally correct, while \EX{I1} and \EX{I2} are
incorrect.

\subsection{Clustering of Correct Student Solutions}\seclabel{CorrectClassification}

The goal of clustering is two-fold.
(1)~\emph{Scalability}: elimination of \emph{dynamically equivalent} correct
solutions that the repair algorithm would otherwise consider separately.
(2)~\emph{Diversity of repairs}: mining of \emph{dynamically equivalent}, but
\emph{syntactically different} expressions from the same cluster, which are
used to repair incorrect student attempts.
We discuss the notion of \emph{dynamic equivalence} next.

\paragraph{Matching}
The clusters in our approach are the \emph{equivalence classes} of a
\emph{matching} relation.
We say that two programs $\prog$ and $\progb$ match, written $\prog \progsim \progb$,
when: (1)~they have the same control-flow (see the discussion below), and (2)~there is a total
bijective relation between the variables of $\prog$ and $\progb$, such that
related variables take the same values, in the same order, during the program
execution on the same inputs.
This is inspired by the notion of a \emph{simulation
  relation}~\cite{sim:milner71}, adapted for a dynamic program analysis:
whereas a simulation relation establishes that a program $\prog$ produces
exactly the same values as program $\progb$ at corresponding program locations
\emph{for all inputs}, we are interested only in a \emph{fixed finite set of
  inputs}.
Therefore, we also call this notion \emph{dynamic} equivalence, to stress that
we use dynamic program analysis.

\paragraph{Control-flow}
Our algorithms consider two control-flows the same if they have the same
\emph{looping structure}.
That is, any loop-free sequence of code is treated as a single block; in
particular, blocks can include (nested) if-then-else statements without loops
(similar to~\citet{large-block}).
We point out that we could have picked a different \emph{granularity} of
control-flow; e.g., to treat only straight line of code (without branching) as
a block.
We have picked this granularity because it enables matching of
programs that have different branching-structure, and as a result our algorithm
is able to generate repairs that involve missing if-then-else statements.

\emph{Example.}
Programs \EX{C1} and \EX{C2}, from \figref{MotivExamples}~(a)~and~(b),
\emph{match}, because:
(1)~they have the same control-flow since there is only a single loop in both
programs;
(2)~there is the bijective variable relation
$\simrel = \{ \mathit{poly} \mapsto \mathit{poly}, \mathit{deriv} \mapsto \mathit{result}, \cond \mapsto \cond, i \mapsto e, \retvar \mapsto \retvar \}$,
where variables $\cond$ and $\retvar$ are \emph{special variables} denoting the
loop condition and the return value, which we need to make the control-flow and
the return values explicit.
For example, on the input $\mathit{poly} = [6.3, 7.6, 12.14]$, the variable
$\mathit{result}$, takes the value $[]$ before the loop, the sequence of values
$[7.6], [7.6,24.28]$ inside the loop, and the value $[7.6,24.28]$ after the
loop.
Exactly the same values are taken by the variable $\mathit{deriv}$; and
similarly for all the other variables.
Therefore, \EX{C1} and \EX{C2} belong to the same cluster, which we
denote by $\classes$.
For the further discussion, we need to fix one correct solution as a
\emph{cluster representative}; we pick \EX{C1}, although it is irrelevant which
program from the cluster we pick.

Although the expressions in the assignments to variables $\mathit{result}$ and
$\mathit{deriv}$ generate the same values (i.e., they match or they are
dynamically equivalent), the assignment expressions inside the loops are
syntactically quite different;
we state the expressions in terms of the variables of the cluster
representative \EX{C1}:
\begin{itemize}
\item $\inlinecode{append(result, float(poly[e]*e))}$ in \EX{C1}, and
\item $\inlinecode{result + [float(e)*poly[e]]}$ in \EX{C2}, where
\end{itemize}
the second expression has been obtained by replacing the variables from \EX{C2}
with variables from \EX{C1} using the variable relation $\simrel$ (e.g., we
have replaced $\mathit{deriv}$ with $\mathit{result}$, because
$\simrel(\mathit{deriv}) = \mathit{result}$).
In our benchmark we found 15 syntactically different, but dynamically
equivalent, ways to write expressions for the assignment to $\mathit{result}$,
and 6 different ways to write the return expression (observing only different
ASTs, and ignoring formatting differences).
Some of these examples are shown in \figref{MotivExamples}~(c)~and~(d),
respectively.  As we discuss in the next section, these different expressions
are used by our repair algorithm.

\subsection{Repair of Incorrect Student Attempts}\seclabel{IncorrectRepair}

\begin{figure}[ht]
  \centering
  \includegraphics[width=0.45\textwidth]{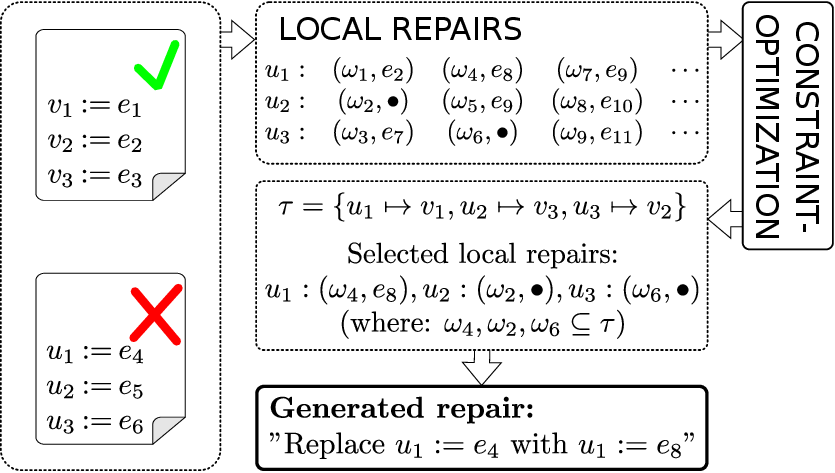} \\
  \caption{High-level overview of the repair algorithm.}
  \figlabel{OverviewRepair}
\end{figure}

Our repair algorithm takes an incorrect student attempt (which we call an
\emph{implementation}) and a \emph{cluster} (of correct programs), and returns
a \emph{repair}; a repair modifies the implementation such that the repaired
implementation and the cluster representative \emph{match}.
The top-level repair procedure takes an implementation and runs the repair
algorithm on each cluster separately.
Each repair has a certain cost w.r.t. some cost metric.
In this paper we use syntactic distance.
Finally, the repair with the minimal cost is chosen.

\figref{MotivExamples}~(e) and~(f) shows two incorrect programs, \EX{I1} and
\EX{I2}, and the generated repairs in~(g) and~(h), respectively.
These repairs were generated using cluster $\classes$, with its representative
\EX{C1}.
Our algorithm also considered other cluster besides $\classes$ (which are not
discussed here), but found the smallest repair using $\classes$.
In the rest of this section we discuss the repair algorithm on a single
cluster.

Our algorithm generates repairs in two steps, which we discuss in more detail
next:
(1)~The algorithm generates a set of \emph{local repairs} for each
implementation variable (its assigned expression).
(2)~Using constraint-optimization techniques, the algorithm selects a
\emph{consistent} subset of the local repairs with the smallest cost.
The high-level overview of the algorithm is given in \figref{OverviewRepair}.

\paragraph{Local Repairs}
A local repair ensures that an implementation expression $\expr_\impl$, either
modified or unmodified, \emph{matches} a corresponding cluster representative
expression $\expr_\classes$.
To establish that the expressions match, we need a variable relation that
translates implementation variables to cluster variables, so that the
expressions range over the same variables.
We point out that for \emph{expression matching} it is sufficient to consider
\emph{partial variable relations} that relate variables of the expressions, as
opposed to \emph{total variable relations} that relate all program variables,
which we considered above for \emph{program matching}.

Let $\var$ be an implementation variable, with expression $\expr_\impl$
assigned to it at some program location.
We say that $(\parmap, \expr_\repaired)$ is a local repair for $\var$, if the
expressions $\expr_\classes$ and $\expr_\repaired$ match, where $\parmap$ is a
partial variable relation that establishes the matching.
In this case the implementation expression $\expr_\impl$ is modified to
the \emph{repaired expression} $\expr_\repaired$.
We discuss below how the algorithm generates the repaired expression
$\expr_\repaired$.
We say that $(\parmap, \nomod)$ is a local repair for $\var$, if the
expressions $\expr_\classes$ and $\expr_\impl$ match, where $\parmap$ is a
partial variable relation that establishes the matching.
In this case the implementation expression $\expr_\impl$ remains unmodified.

We illustrate the notion of local repairs on \EX{I1} with regard to the cluster
representative \EX{C1}:
\begin{itemize}
\item[1.] $(\parmap_1, \nomod)$ is a local repair for $\mvar{new}$ before the loop,
  where $\parmap_1 = \{ \mvar{new} \mapsto \mvar{result} \}$, since the
  expressions of $\mvar{new}$ and $\mvar{result}$ match (i.e., they take the
  same values).
\item[2.] $(\parmap_2, \inlinecode{if new==[]: return [0.0] else:}$ \\
  $\inlinecode{return new})$ is a local repair for $\mvar{return}$
  after the loop, where $\parmap_2 = \{ \mvar{new} \mapsto \mvar{result}, \mvar{return} \mapsto \mvar{return} \}$,
  since then the return expressions match.
\item[3.] $(\parmap_3, \inlinecode{if poly==[]: return [0.0] else:}$ \\
  $\inlinecode{return poly})$ is a local repair for $\mvar{return}$
  after the loop, where $\parmap_3 = \{ \mvar{poly} \mapsto \mvar{result}, \mvar{return} \mapsto \mvar{return} \}$,
  since then the return expressions match.
\end{itemize}
The algorithm generates a set of such local repairs for each variable
and each program location in the implementation.

\paragraph{Finding a repair}
A (whole program) repair is a \emph{consistent subset} of the generated local
repairs, such that there is \emph{exactly one local repair for each variable and each
program location in the implementation}.
A set of local repairs is consistent when all partial variable relations in the
local repairs are subsets of some total variable relation.
For example, local repairs (1) and (3) from above are \emph{inconsistent},
since we have $\parmap_1(\mvar{new}) = \mvar{result}$ and
$\parmap_3(\mvar{poly}) = \mvar{result}$, and hence there is no total variable
relation that is consistent with both $\parmap_1$ and $\parmap_3$.
On the other hand, local repairs (1) and (2) are consistent, since there is a total variable relation consistent both with
$\parmap_1$ and $\parmap_2$:
$\{ \mvar{poly} \mapsto \mvar{poly}, \mvar{new} \mapsto \mvar{result}, i \mapsto e, \cond \mapsto \cond, \mvar{return} \mapsto \mvar{return} \}$.

There are many choices for a whole program repair, that is, choices for a
consistent subset of the generated local repairs;
however, we are interested in one that has the smallest cost.
Our algorithm finds such a repair using constraint-optimization techniques.

\paragraph{Generating the set of potential local repairs}
The repair algorithm takes expressions from the different correct solutions in
order to generate the set of potential local repairs ($\expr_\repaired$
discussed above).
The algorithm translates these expressions to range over the implementation
variables, using a partial variable relation.
Since each cluster expression matches a corresponding cluster representative
expression (recall the discussion in the previous sub-section), this partial
variable relation ensures that $\expr_\repaired$ matches a corresponding
cluster representative expression (i.e., that it is a correct repair).

For example, the generated repair for \textbf{I2}, shown in
\figref{MotivExamples}~(h), combines the following expressions from the cluster:
\begin{itemize}
\item The first modification is generated using the expression
  $\inlinecode{range(1, len(poly))}$
  from \textbf{C1}, at line 3, using the partial variable relation $\{ \mvar{poly} \mapsto \mvar{poly} \}$.
\item The second modification is generated using the expression (4.) from \figref{MotivExamples}~(c),
  using the partial variable relation $\{ \mvar{result} \mapsto \mvar{result}, \mvar{poly} \mapsto \mvar{poly}, i \mapsto e  \}$.
\item The third modification is generated using the expression (3.) from \figref{MotivExamples}~(d),
  using the partial variable relation $\{ \retvar \mapsto \retvar, \mvar{result} \mapsto \mvar{result} \}$.
\end{itemize}

%% file: model.tex
\section{Program Model}\seclabel{model}

In this section we define a program model that captures key aspects
of imperative languages (e.g., \clang, \pylang).
This model allows us to formalize our notions of program matching and program repair.

\begin{definition}[Expressions]\deflabel{exprs}
  Let $\vars$ be a set of variables, $\consts$ a set of constants,
  and $\ops$ a set of operations.
  The set of expressions $\exprs$ is built from variables, constants and operations in the usual way.
  We fix a set of \emph{special variables} $\specvars{\vars} \subseteq \vars$.
  We assume that $\specvars{\vars}$ includes at least the variable $\cond$, which we will use
  to model conditions, and the variable $\retvar$, which we will use to model return values.
\end{definition}

\defref{exprs} can be instantiated by a concrete programming language.
For example, for the \clang language, $\consts$ can be chosen to be the set of
all \clang constants (e.g., integer, float), and $\ops$ can be chosen to be the
set of unary and binary \clang operations as well as library functions.
The special variables are assumed to not appear in the original program text,
and are only used for modelling purposes.

\begin{definition}[Program]\deflabel{progs}
  A \emph{program} $\prog = (\locs_\prog, \linit,\allowbreak \vars_\prog, \blabel_\prog, \allowbreak \succfun_\prog)$ is a tuple,
  where $\locs_\prog$ is a (finite) set of \emph{locations}, $\linit \in \locs_\prog$
  is the \emph{initial location}, $\vars_\prog$ is a (finite) set of \emph{program variables},
  $\blabel_\prog : (\locs_\prog \times \vars_\prog) \to \exprs$
  is the \emph{variable update function} that assigns an expression to every location-variable pair,
  and $\succfun_\prog : (\locs_\prog \times \{ \true, \false \}) \to (\locs_\prog \cup \{ \lend \})$
  is the \emph{successor function},
  which either returns a successor location in $\locs_\prog$ or the special value $\lend$ (we assume $\lend \not\in \locs_\prog$).
  We drop the subscript $\prog$ when it is clear from the context.
\end{definition}

We point the reader to the discussion around the semantics below for an
intuitive explanation of the program model.

\begin{definition}[Computation domain, Memory]\deflabel{compdom-mem}
  We assume some (possibly infinite) set $\compdom$ of values, which we call the {\em computation domain},
  containing at least the following values:
  (1) $\true$ ({\em bool true});
  (2) $\false$ ({\em bool false}); and
  (3) $\undeff$ ({\em undefined}).

  Let $\vars$ be a set of variables.
  A {\em memory} $\mem : (\vars \cup \vars') \to \compdom$ is a mapping from program
  variables to values, where the set $\vars' = \{ \var' \mid \var \in \vars \}$
  denotes the primed version of the variables in $\vars$;
  let $\mems_\vars$ denote the set of all memories over variables $\vars \cup \vars'$.
\end{definition}

Intuitively, the primed variables are used to denote the variable values after
a statement has been executed (see the discussion around the semantics below).

\begin{definition}[Evaluation]\deflabel{evalfun}
  A function $\semfun{\cdot} : \exprs \to \mems \to \compdom$ is
  an {\em expression evaluation function},
  where $\semfun{\expr}(\mem) = \compval$ denotes that $\expr$, on a memory $\mem$,
  evaluates to a value $\compval$.
\end{definition}

The function $\semfun{\cdot}$ is defined by a concrete programming language.
The function returns the undefined value ($\undeff$) when an
error occurs during the execution of an actual program.

\begin{definition}[Program Semantics]\deflabel{progsem}
  Let $\prog = (\locs_\prog, \linit,\allowbreak \vars_\prog, \allowbreak \blabel_\prog, \succfun_\prog)$ be
  a program.
  A sequence of location-memory pairs $\trace \in (\locs_\prog \times \mems_{\vars_\prog})^*$ is called a {\em trace}.
  Given some ({\em input}) memory $\inmem$, we write $\semfun{\prog}(\inmem) = (\loc_1,\mem_1) \cdots (\loc_n,\mem_n)$
  if:
  (1) $\loc_1 = \linit$;
  (2) $\mem_1(\var) = \inmem(\var)$ for all $\var \in \vars_\prog$;
  (3) 
  (a) $\mem_j(\var') = \semfun{\blabel_\prog(\loc_j, \var)}(\mem_j)$, and 
  (b) $\mem_{j+1}(\var) = \mem_{j}(\var')$, and
  (c) $\loc_{j+1} = \succfun_\prog(\loc_j, \mem_j(\cond'))$, for all $\var \in \vars_\prog$ and $0 \leq j < n$; and
  (4) $\succfun_\prog(\loc_n, b) = \lend$, for any $b \in \{ \true, \false \}$.
\end{definition}

For some trace element $(\loc,\mem) \in \trace$ and a variable $\var$,
$\mem(\var)$ denotes the value of $\var$ before the location $\loc$ is evaluated (the \emph{old value} of $\var$ at $\loc$),
and $\mem(\var')$ denotes the value of $\var$ after the location $\loc$ is evaluated (the \emph{new value} of $\var$ at $\loc$).
The definition of $\semfun{\prog}(\inmem)$ then says:
\begin{itemize}
\item[(1)] The first location of the trace is the initial location $\linit$.
\item[(2)] The \emph{old values} of the variables at the initial location $\linit$ are defined by the input memory $\inmem$.
\item[(3a)] The \emph{new value} of variable $\var$ at location  $\loc_j$ is determined
by the semantic function $\semfun{\cdot}$ evaluated on the expression $\blabel_\prog(\loc_j,\var)$.
\item[(3b)] The old value of variable $\var$ at location $\loc_{j+1}$ is
 equal to the new value at location $\loc_j$.
\item[(3c)] The next location $\loc_{j+1}$ in a trace is determined by the successor function $\succfun_\prog$ for
the current location $\loc_j$ and the new value of $\cond$ at $\loc_j$ (i.e., $\mem_j(\cond')$).
\item[(4)]~The successor of the last location, $\loc_n$, for any Boolean $b \in \{ \true, \false \}$, is the
end location $\lend$.
\end{itemize}

\paragraph{Modelling of if-then-else statements}
In our implementation we model if-then-else statements differently, according to when they
contain loops and when they are loop-free
(as mentioned in \secref{CorrectClassification}).
In the former case, the branching is modelled, as usual, directly in the
control-flow.
In the latter case, (loop-free) statements are (recursively) converted to
$\inlinecode{ite}$ expressions that behave like a \clang ternary operator
(as in the example that follows).

\emph{Example.}
We now show how a concrete program (\EX{C1} from \figref{MotivExamples}~(a))
is represented in our model.
The set of locations is $\locs = \{ \lpre,\lcond,\lloop,\lpost \}$,
where $\linit = \lpre$ is the location before the loop, and the initial location,
$\lcond$ is the location of the loop condition, $\lloop$ is the loop body,
and $\lpost$ is the location after the loop.
The successor function is given by $\succfun(\lpre,\true)=\succfun(\lpre,\false)=\lcond$,
$\succfun(\lcond,\true) = \lloop$, $\succfun(\lcond,\false) = \lpost$,
$\succfun(\lloop, \true)=\succfun(\lloop, \false)$ $=\lcond$,
and $\succfun(\lpost,\true)=\succfun(\lpost,\true)=\lend$.
Note that for non-branching locations the successor functions points to the
same location for both $\true$ and $\false$.

The set of variables is $\vars = \{ \mvar{poly}, \mvar{result}, \mvar{e}, \retvar, \cond, \mvar{it} \}$,
where $\mvar{it}$ is used to model \pylang's {\em for-loop iterator}.
An iterator is a sequence whose elements are assigned, one by one, to some variable ($e$ in this example) in each loop iteration.
The expression labeling function is given by:
\begin{itemize}
\item $\blabel(\lpre,\mvar{result}) = \inlinecode{[]}$,
\item $\blabel(\lpre,\mvar{it}) = \inlinecode{range(1,len(poly))}$,
\item $\blabel(\lcond,\cond) = \inlinecode{len(it)>0}$,
\item $\blabel(\lloop,e) = \inlinecode{ListHead(it)}$,
\item $\blabel(\lloop,\mvar{it}) = \inlinecode{ListTail(it)}$,
\item $\blabel(\lloop,\mvar{result}) = \inlinecode{append(float(poly[e]*e))}$,
\item $\blabel(\lpost,\retvar) = \inlinecode{ite(result==[], [0.0], result)}$.
\end{itemize}
For any variable $\var$ that is unassigned at some location $\loc$ we set
$\blabel(\loc,\var) = \var$, i.e., the variable remains unchanged.

Finally, we state the trace when {\bf C1} is executed on $\inmem = \{ \mvar{poly} \pts [6.3, 7.6, 12.14] \}$.
We state only defined variables that change from one trace element to the next.
Otherwise, we assume the values remain the same or are undefined ($\undeff$) (if no previous value existed).
$\semfun{\text{\bf C1}}(\inmem) =$
$(\lpre, \{ \mvar{poly} \pts [6.3,7.6,12.14], \mvar{result'}=[], i'=0, \mvar{it}'=[1,2] \})$,
$(\lcond, \{  \cond'=\true \})$,
$(\lloop, \{ e' \pts 1, \mvar{it}'=[2], i' \pts 1, \mvar{result'} \pts [7.6] \}$,
$(\lcond, \{  \cond'=\true \})$,
$(\lloop, \{ e' \pts 2, \mvar{it}'=[], i \pts 3, \mvar{result'} \pts [7.6, 24.28] \}$,
$(\lcond, \{  \cond'=\false \})$,
$(\lpost,$ $\{  \retvar$ $\pts$ \\
$[7.6,24.28] \})$.

%% file: matching.tex
\section{Matching and Clustering}\seclabel{clustering}

In this section we formally define our notion of \emph{matching}.

Informally, two programs match, if (1)~the programs have the \emph{same
  control-flow} (i.e., the same looping structure), and (2)~the corresponding
variables in the programs take the same values in the same order.
For the following discussion we fix two programs,
$\prog=(\locs_\prog, \loc_{\init_\prog}, \vars_\prog, \blabel_\prog, \succfun_\prog)$
and $\progb=(\locs_\progb, \loc_{\init_\progb}, \vars_\progb, \blabel_\progb, \succfun_\progb)$.

\begin{definition}[Program Structure]\deflabel{progstruct}
Programs $\prog$ and $\progb$ have the \emph{same control-flow}
if there exists a bijective function, called \emph{structural matching}, $\isom : \locs_\progb \to \locs_\prog$, s.t.,
for all $\loc \in \locs_\progb$ and $b \in \{ \true, \false \}$,
$\succfun_\prog(\isom(\loc), b) = \isom(\succfun_\progb(\loc,b))$.
\end{definition}

We remind the reader, as discussed in \secref{overview} and \secref{model},
that we encode any loop-free program part as single control-flow location;
as a result we compare only the \emph{looping structure} of two programs.

We note that both our matching and repair algorithms require the existence of a
\emph{structural matching} $\isom$ between programs.
Therefore, in the rest of the paper we assume that such a $\isom$ exists between any two programs
that we discuss, and assume
that $\locs = \locs_\prog = \locs_\progb$ and $\succfun = \succfun_\prog = \succfun_\progb$,
since they can be always converted back and forth using $\isom$.
Next, we state two definitions that will be useful later on.
\begin{definition}[Variables of expression]
  Let $\expr$ be some expression, by $\exprvars(\expr) = \{ \var~\mid~\var \in \expr \}$
  we denote the \emph{set of variables used} in the expresison $\expr$.
  We also say that $\expr$ \emph{ranges over} $\exprvars(\expr)$.
\end{definition}
\begin{definition}[Variable substitution]\deflabel{translation}
  Let $\simrel : \vars_1 \to \vars_2$ be some function that maps variables $\vars_1$ to variables $\vars_2$.
  Given an expression $\expr$ over variables $\vars_1$, i.e., $\exprvars(\expr) \subseteq \vars_1$,
  by $\simrel(\expr)$ we denote the expression,
  which we obtain from $\expr$, by substituting $\var$ with $\simrel(\var)$ for all $\var \in \vars_1$.
  Note that $\exprvars(\simrel(\expr)) \subseteq \vars_2$.

  Given a memory $\mem \in \mems_{\vars_1}$, over variables $\vars_1$, we define
  $\simrel(\mem) = \{ \simrel(\var) \pts \mem(\var), \simrel(\var)' \pts \mem(\var') \mid \var \in \vars_1 \}$,
  that is, the memory where $\var_1$ is substituted with $\var_2$, for all $\var_2 = \simrel(\var_1)$.
  We lift substitution to a trace $\trace = (\loc_1,\mem_1) \cdots (\loc_n,\mem_n) \in (\locs \times \mems_{\vars_1})^{*}$, by applying it
  to the each element: $\simrel(\trace) = (\loc_1,\simrel(\mem_1)) \cdots \allowbreak (\loc_n,\simrel(\mem_n)) \in (\locs \times \mems_{\vars_2})^{*}$.
\end{definition}

In the rest of the paper
we call a \emph{bijective function} $\simrel : \vars_1 \to \vars_2$, between two sets of variables $\vars_1$ and $\vars_2$,
a \textbf{total variable relation} between $\vars_1$ and $\vars_2$.
Next we give a formal definition of matching between two programs.
Afterwards, we give a formal definition of matching between two expressions.
The definitions involve execution of the programs on a set of inputs.
\begin{definition}[Program Matching]\deflabel{matching}
  Let $\ins$ be a set of inputs,
  and let $\trace_{\prog,\inmem} = \semfun{\prog}(\inmem)$ and $\trace_{\progb,\inmem} = \semfun{\progb}(\inmem)$
  be sets of traces obtained by executing $\prog$ and $\progb$ on $\inmem \in \ins$, respectively.

  We say that $\prog$ and $\progb$ \emph{match} over a set of inputs $\ins$,
  denoted by $\prog \sim_\ins \progb$, if there exists a total variable relation
  $\simrel : \vars_\progb \to \vars_\prog$, such that $\trace_{\prog,\inmem} = \simrel(\trace_{\progb,\inmem})$,
  for all inputs $\inmem \in \ins$.
  We call $\simrel$ a \emph{matching witness}.
\end{definition}
Intuitively, a matching witness $\simrel$ defines a way of translating $\progb$
to range over variables $\vars_\prog$, such that $\prog$ and $\progb$
translated with $\simrel$ produce the same traces.

Given a set of inputs $\ins$, $\progsim_\ins$ is an
{\em equivalence relation} on a set of programs $\specs$:
the \emph{identity} relation on program variables gives a matching witness for \emph{reflexivity},
the \emph{inverse} $\simrel^{-1}$ of some total variable relation $\simrel$ gives
a matching witness for \emph{symmetry},
and the \emph{composition}  $\simrel_1 \circ \simrel_2$ of some total variables relations $\simrel_1,\simrel_2$
gives a matching witness for \emph{transitivity}.

\begin{figure}[t]
  \centering
  \includecode{match-algo.txt}{0.26} 
  \caption{The Matching Algorithm.}
  \figlabel{MatchAlgorithm}
\end{figure}

The algorithm for finding $\simrel$ is given in~\figref{MatchAlgorithm}:
given two programs $\prog$ and $\progb$ and a set of inputs $\ins$,
it returns a matching witness $\simrel$, if one exists.
The algorithm first executes both programs on the inputs (lines 3 and 4),
 and then for each variable $\var_2 \in \vars_\progb$ finds a set of
variables from $\vars_\prog$ that take the same values during
execution, thus defining a set of potential matches $M \subseteq \vars_\progb \times \vars_\prog$ (lines 5-10);
here $\trace|_\var$, denotes a projection of values of variable $\var$ from a trace $\trace$, that is:
$((\loc_1,\mem_1) \cdots (\loc_n,\mem_n))|_\var = \mem_1(\var') \cdots \mem_n(\var')$.
The matching witness is then a bijective mapping $\simrel \subseteq M$, if one exists (line 11);
$\simrel$ can be found in $M$ by constructing a \emph{maximum bipartite
  matching} in the \emph{bipartite graph} defined by $M$.
We note that this problem can be solved in polynomial time, w.r.t. the number
of the edges and vertices in $M$~(e.g.,\citet{Takeaki1997}).

\begin{definition}[Expression matching]\deflabel{expr-match}
  Let $\traces \subseteq (\locs \times \mems_{\vars_\prog})^{*}$
  be a set of traces over variables $\vars_\prog$, and
  let $\expr_1$ and $\expr_2$ be two expressions over variables $\vars_\prog$,
  at some location $\loc \in \locs$.

  We say that $\expr_1$ and $\expr_2$ \emph{match} over a set of traces $\traces$,
  denoted $\expr_1 \esim_{\traces,\loc} \expr_2$, if
  $\semfun{\expr_1}(\mem) = \semfun{\expr_2}(\mem)$,
  for all $(\mem, \loc_i) \in \trace$ where $\loc_i = \loc$,
  and all $\trace \in \traces$.
\end{definition}

Expression matching says that two expressions produce the same values,
when considering the memories at location $\loc$, in a set of traces $\traces$.
In the following lemma we state that expression matching is equivalent to
program matching; this lemma will be useful for our repair algorithm, which we will state in the next section.

\begin{lemma}[Matching Equivalence]\lemlabel{match-equiv}
  Let $\ins$ be a set of inputs, and let $\traces_\ins = \{ \semfun{\prog}(\inmem) \mid \inmem \in \ins \}$ be
  a set of traces obtained by executing $\prog$ on $\ins$.
  We have the following equivalence:
  $\prog \sim_\ins \progb$ witnessed by $\simrel : \vars_\progb \to \vars_\prog$, if and only if,
  $\expr_\prog \esim_{\traces_\ins,\loc} \simrel(\expr_\progb)$, for all $(\loc,\var_1) \in \locs \times \vars_\prog$,
  where $\var_1 = \simrel(\var_2)$, $\expr_\prog = \blabel_\prog(\loc, \var_1)$, and
  $\expr_\progb = \blabel_\progb(\loc, \var_2)$.
\end{lemma}
\begin{proof}
  ``$\Rightarrow$'': Directly from the definitions.
  ``$\Leftarrow$'': By induction on the length of the trace $\trace = \semfun{\prog}(\inmem)$ for some $\inmem \in \ins$.
\end{proof}

\paragraph{Clustering}
We define clusters as the equivalence classes of $\sim_\ins$.
For the purpose of matching and repair we pick an arbitrary class
representative from the class and collect expressions from all programs in the same cluster:

\begin{definition}[Cluster]
  Let $\specs$ be a set of (correct) programs.
  A {\em cluster} $\classes \subseteq \specs$ is an \emph{equivalence class}
  of $\progsim_\ins$.
  Given some cluster $\classes$, we fix some \emph{arbitrary class representative} $\prog_\classes \in \classes$.

  We define the set $\cexprs_\classes(\loc, \var_1)$ of \emph{cluster expressions} for a pair $(\loc, \var_1) \in \locs \times \vars_{\prog_\classes}$:
  $\expr_1 \in \cexprs_\classes(\loc, \var_1)$ iff there is some
  $\progb \in \classes$ witnessed by $\simrel : \vars_\progb \to \vars_{\prog_\classes}$
  such that $\var_1 = \simrel(\var_2)$, $\expr_2 = \blabel_\progb(\loc, \var_2)$ and $\expr_1 = \simrel(\expr_2)$.
\end{definition}

Note that it is irrelevant which program from $\classes$ is chosen as cluster representative $\prog_\classes$;
we just need to fix some program in order to be able to define the expressions $\cexprs_\classes$ over a common set of variables $\vars_{\prog_\classes}$.
We note that by definition the sets of expressions $\cexprs_\classes$ have the following property:
for all $\expr_1,\expr_2 \in \cexprs_\classes(\loc,\var)$ we have $\expr_1 \esim_{\traces_\ins,\loc} \expr_2$ that is,
expressions $\expr_1$ and $\expr_2$ match.

\emph{Example.}
In \secref{CorrectClassification} we discussed why the solutions \EX{C1}
and \textbf{C2} match; therefore these two solutions belong to the same
cluster, which we denote here by $\classes$, and chose \EX{C1} as its representative $\prog_\classes$.
Also, in \figref{MotivExamples}~(c)~and~(d) we gave examples of different
equivalent expressions of assignment to the variable $\mathit{result}$ inside
the loop body and the return statement after the loop, respectively.
To be more precise, these were examples of sets
$\cexprs_{\classes}(\lloop, \mathit{result})$ and
$\cexprs_{\classes}(\lpost,\mathit{return})$, respectively.

%% file: repair.tex
\section{Repair Algorithm}\seclabel{repair}

In the previous section we defined the notion of a matching between two
programs.
In this section we consider an implementation $\pimpl$ and a cluster $\classes$
(with its representative $\prog_\classes$) between which there is no matching.
We assume that $\pimpl$ and $\prog_\classes$ have the same control-flow.
The goal is to \emph{repair} $\pimpl$; that is, to modify $\pimpl$ minimally,
w.r.t. some notion of cost, such that the repaired program matches the cluster.
More precisely, the repair algorithm searches for a program $\prog_\repaired$,
such that $\prog_\classes \sim_\ins \prog_\repaired$, and $\prog_\repaired$ should be syntactically close to $\prog_\impl$.
Therefore, our repair algorithm can be seen as a \emph{generalization of
  clustering to incorrect programs}.

We first define a version of the repair algorithm that does not change the set of variables, i.e., $\vars_\impl = \vars_{\repaired}$.
Below we extend this algorithm to include changes of variables, i.e., we allow $\vars_\impl \neq \vars_{\repaired}$.
In both cases the control-flow of $\pimpl$ remains the same.

For the following discussion we fix some set of inputs~$\ins$.
Let $\traces = \{ \semfun{\prog_\classes}(\inmem) \mid \inmem \in \ins \}$ be the set of traces
of cluster representative $\prog_\classes$ for the inputs $\ins$.

As we discussed in the previous section, two programs match if all of their
corresponding expressions match (see \lemref{match-equiv}).
Therefore the key idea of our repair algorithm is to consider a set of
\emph{local repairs} that modify individual implementation expressions.
We first discuss local repairs; later on we will discuss how to combine local repairs into a full program repair.

Local repairs are defined with regard to \emph{partial variable relations}.
It is enough to consider partial variable relations (as opposed to the total
variable relations needed for matchings) because these relations only need to
be defined for the expressions that need to be repaired.

\begin{definition}[Local Repair]
  Let $(\loc,\var_2) \in \locs \times \vars_\impl$ be a location-variable pair
  from $\pimpl$, and let $\expr_\impl = \blabel_\impl(\loc,\var_2)$ be the
  corresponding expression.
  Further, let $\parmap : \vars_\impl \parto \vars_{\prog_\classes}$ be a
  \emph{partial variable relation} such that $\var_2 \in \dom(\parmap)$, let
  $\var_1 = \parmap(\var_2)$ be the related cluster representative
  variable, and let $\expr_{\classes} = \blabel_{\prog_\classes}(\loc, \var_1)$ be the
  corresponding expression.

  A pair $\repair = (\parmap, \expr_\repaired)$, where $\expr_\repaired$ is an expression over implementation variables $\vars_\impl$,
  is a \emph{local repair} for $(\loc,\var_2)$ when
  $\expr_\classes \esim_{\traces,\loc} \parmap(\expr_\repaired)$ and $\exprvars(\expr_\repaired) \subseteq \dom(\parmap)$.
  A pair $\repair = (\parmap, \nomod)$ is a \emph{local repair} for $(\loc,\var_2)$ when
  $\expr_\classes \esim_{\traces,\loc} \parmap(\expr_\impl)$ and $\exprvars(\expr_\impl) \subseteq \dom(\parmap)$.

  We define the cost of a local repair $\repair = (\parmap, \expr_\repaired)$ as $\mvar{cost}(r) = \mvar{diff}(\expr_\impl, \expr_\repaired)$ and the cost of a local repair $\repair = (\parmap, \nomod)$ as $\mvar{cost}(r) = 0$.
\end{definition}

We comment on the definition of a local repair.
Let $(\loc,\var_2) \in \locs \times \vars_\impl$ be a location-variable pair from $\pimpl$, let $\expr_\impl = \blabel_\impl(\loc,\var_2)$ be the  corresponding expression, and let $\repair$ be a local repair for some $(\loc,\var_2)$.
In case $\repair = (\parmap, \nomod)$, the expression $\expr_\impl$ matches the corresponding expression of the cluster representative under the partial variable mapping $\parmap$; this repair has cost zero because the expression $\expr_\impl$ is not modified.
In case $\repair = (\parmap, \expr_\repaired)$, the expression $\expr_\repaired$ constitutes a modification of $\expr_\impl$ that matches the corresponding expression of the cluster representative under the partial variable mapping $\parmap$; this repair has some cost $\mvar{diff}(\expr_\impl, \expr_\repaired)$.
In our implementation we define $\mvar{diff}(\expr_\impl, \expr_\repaired)$ to be the tree edit distance~\cite{Tai:1979,Zhang:1989}
between the abstract syntax trees (ASTs) of the expressions $\expr_\impl$ and
$\expr_\repaired$.

\emph{Example.}
We remind the reader that in \secref{IncorrectRepair} we discussed three
examples of local repairs for \EX{I1} (\figref{MotivExamples}).
Formally, example~(1) is a local repair for $(\loc_1, \mvar{new})$, while~(2)
and~(3) are local repairs for $(\loc_4, \retvar)$.
Next we state how to combine local repairs into a full program repair.

\begin{definition}[Repair]
  Let $R$ be a function that assigns to each pair $(\loc,\var) \in \locs \times \vars_\impl$ a \emph{local repair} for $(\loc,\var)$.
  We say that $R$ is \emph{consistent}, if there exists a total variable
  relation $\simrel : \vars_\impl \to \vars_{\prog_\classes}$, such that
  $\parmap \subseteq \simrel$, for all $R(\loc,\var) = (\parmap, -)$.
  A consistent $R$ is called a \emph{repair}.
  We define the \emph{cost} of $R$ as the sum of the costs of all local repairs:
  $\mvar{cost}(R) = \sum_{(\loc,\var) \in \locs \times \vars_\impl} \mvar{cost}(R(\loc,\var))$.

  A repair $R$ defines a \emph{repaired implementation} $\prog_\repaired = (\locs, \linit, \vars_\impl, \blabel_\repaired, \succfun)$,
  where $\blabel_\repaired(\loc, \var) = \expr_\repaired$ if $M(\loc,\var) = (-, \expr_\repaired)$,
  and $\blabel_\repaired(\loc,\var) = \blabel_\impl(\loc,\var)$ if $M(\loc, \var) = (-, \nomod)$,
  for all $(\loc,\var) \in \locs \times \vars_\impl$.
\end{definition}

\begin{theorem}[Soundness of Repairs]
  $\prog_\classes \sim_{\ins} \prog_\repaired$.
\end{theorem}
\begin{proof}(sketch)
  From the definition of $R(\loc, \var_2)$, we have $\expr_\classes \esim_{\traces,\loc} \simrel(\expr_\repaired)$,
  for all $(\loc, \var_2) \in \locs \times \vars_\impl$,
  where $\var_1 = \simrel(\var_2)$, $\expr_\classes = \blabel_\classes(\loc, \var_1)$ and $\expr_\repaired = \blabel_\repaired(\loc, \var_2)$.
  Then it follows from \lemref{match-equiv} that $\prog_\classes \sim_\ins \prog_\repaired$.
\end{proof}

In the above definition we use notation $r = (\parmap, -)$ when $\expr_\repaired$ or $\nomod$ is not important in $r$;
similarly we use $r = (-, \expr_\repaired)$ and $r = (-, \nomod)$ when $\parmap$ is not important in $r$.

\emph{Example.}
The repair for example \textbf{I1} (\figref{MotivExamples}) corresponds to the
total variable relation
$\simrel = \{ \mathit{poly} \mapsto \mathit{poly}, \mathit{new} \mapsto \mathit{result}, \mathit{e} \mapsto i, \mathit{return} \mapsto \mathit{return}, \cond \mapsto \cond \}$.
The repair $M$ includes local repairs (1) and (2) from the
previous examples, where only (2) has cost $> 0$ (see the repair in \figref{MotivExamples}~(g)).

Next we discuss the algorithm for finding a repair.

\begin{figure}[t]
  \centering
  \includecode{repair-algo.txt}{0.42}
  \caption{The Repair Algorithm.}
  \figlabel{RepairAlgorithm}
\end{figure}

\paragraph{The repair algorithm}
The algorithm is given in \figref{RepairAlgorithm}: given
a cluster $\classes$, an implementation $\pimpl$, and a
set of inputs $\ins$, it returns a repair $R$.
The algorithm has two main parts:
First, the algorithm constructs a set of \emph{possible local repairs}; we
define and discuss the possible local repairs below.
Second, the algorithm searches for a consistent subset of the possible local
repairs, which has minimal cost; this search corresponds to solving a
constraint-optimization system.

\begin{definition}[Set of possible local repairs]
\deflabel{possible-local-repair}
  For all $(\loc,\var) \in \locs \times \vars_\impl$, we define the set of \emph{possible local repairs} $\PM(\loc,\var)$ as:
  (1)~$(\parmap, \expr) \in \PM(\loc,\var)$, if $\parmap(\expr) \in \cexprs_\classes(\loc, \parmap(\var))$; and
  (2)~$(\parmap, \nomod) \in \PM(\loc,\var)$, if $\expr_\classes \esim_{\traces,\loc} \parmap(\expr_\impl)$,
  where $\expr_\classes = \blabel_{\prog_\classes}(\loc, \parmap(\var))$ and $\expr_\impl = \blabel_\impl(\loc,\var)$.
\end{definition}

The set of possible local repairs $\PM(\loc,\var)$ includes all expressions
from the cluster $\cexprs_\classes(\loc, \parmap(\var))$, translated by some partial
variable relation $\parmap$ in order to range over implementation variables.
It also includes $(\parmap, \nomod)$ if $\expr_\impl$ matches $\expr_\classes$ under partial variable mapping $\parmap$.
Next we describe how the algorithm constructs the set $\PM(\loc, \var)$ (at lines
4-14).

For the following discussion we fix a pair
$(\loc,\var_2) \in \locs \times \vars_\impl$ (corresponding to line 4),
and some $\var_1 \in \vars_{\prog_\classes}$ (corresponding to line 7);
we set $\expr_\impl = \blabel_\impl(\loc, \var_2)$ and $\expr_\classes = \blabel_{\prog_\classes}(\loc, \var_1)$.
Possible local repairs for $(\loc,\var_2)$ are constructed in two steps:
In the first step, the algorithm checks if there are
partial variable relations $\parmap : \vars_\impl \parto \vars_{\prog_\classes}$ s.t. $\expr_\classes \esim_{\traces,\loc} \parmap(\expr_\impl)$  (at line 9),
and in that case adds a pair $(\parmap, \nomod)$ to $\PM(\loc,\var_2)$ (at line 11).
In the second step, the algorithm iterates over all cluster expressions $\expr = \cexprs_{\prog_\classes}(\loc,\var_1)$ (at line 12),
and all partial variable relations $\parmap : \vars_{\prog_\classes} \parto \vars_\impl$ (at line 13),
and then adds a pair $(\parmap^{-1}, \parmap(\expr))$ to $\PM(\loc,\var_2)$ (at line 14).
We note that
$\parmap^{-1}(\parmap(\expr)) = \expr \in \cexprs_\classes(\loc, \var_1) = \cexprs_\classes(\loc, \parmap^{-1}(\var_2))$,
and thus $(\parmap^{-1}, \parmap(\expr))$ is a possible local repair as in \defref{possible-local-repair}.

We remark that in both steps, the algorithm iterates
over all possible variable relations $\parmap$.
However, since $\parmap$ relates only the variables of a single expression --- 
usually a small subset of all program variables, this iteration is feasible.

\paragraph{Finding a repair with the smallest cost}
Finally, the algorithm uses sub-routine $\textsc{FindRepair}$ (at line 15)
that, given a set of possible local repairs $\PM$, finds a repair with smallest
cost.
\textsc{FindRepair} encodes this problem as a \emph{Zero-One Integer Linear
  Program (ILP)}, and then hands it to an off-the-shelf ILP solver.
Next, we define the ILP problem, describe how we encode the problem of finding
a repair as an ILP problem, and how we decode the ILP solution to a repair.

\begin{definition}[(Zero-One) ILP]
An ILP problem, over variables $\ints = \{ x_1, \dots, x_n \}$, is defined by
an {\em objective function} $\objective = \boxdot \sum_{1 \leq i \leq n} w_i \cdot x_i$, and
a set of {\em linear (in)equalities}  $\constrs$, of the form $\sum_{1 \leq i \leq n} a_i \cdot x_i \trianglerighteq b$.
Where $\boxdot \in \{ \min, \max \}$ and $\trianglerighteq = \{ \geq, = \}$.
A solution to the ILP problem is a variable assignment $\assgn: \ints \to \{0, 1\}$, such that
all (in)equalities hold, and the value of the objective functions is minimal (resp. maximal) for $\assgn$.
\end{definition}

We encode the problem of finding a consistent subset of possible local repairs as an ILP problem with variables
$\ints = \{ x_{\var_1\var_2} \mid \var_1 \in \vars_{\prog_\classes} \text{ and } \var_2 \in \vars_\impl \} \cup \{ x_\repair \mid \repair \in \PM(\loc,\var), (\loc,\var) \in \locs \times \vars_\impl \}$;
that is, one variable for each pair of variables $(\var_1,\var_2)$, and one variable for each possible local repair $\repair$.
The set of constraints $\constrs$ is defined as follows:
\begin{align}
  \left( \textstyle \sum_{\var_2 \in \vars_\impl} x_{\var_1\var_2} \right) = 1 & \text{ for each } \var_1 \in \vars_{\prog_\classes} \\
  \left( \textstyle \sum_{\var_1 \in \vars_{\prog_\classes}} x_{\var_1\var_2} \right) = 1 & \text{ for each } \var_2 \in \vars_\impl \\
  \left( \textstyle \sum_{r \in \PM(\loc,\var)} x_r \right) = 1 & \text { for each } (\loc,\var) \in \locs \times \vars_\impl \\
  -x_r + x_{u_1u_2} \geq 0 & \text{ for each } \repair = (\parmap, -) \in \PM \\ \notag
  & \quad\text{ and each } \parmap(u_2)=u_1
\end{align}
Intuitively, the constraints encode:
\begin{enumerate}
\item Each $\var_1 \in \vars_{\prog_\classes}$ is related to {\em exactly one} of
$\var_2 \in \vars_\impl$.
\item Each $\var_2 \in \vars_\impl$ is related to {\em exactly one} of $\var_1 \in \vars_{\prog_\classes}$.
Together (1) and (2) encode that there is a total variable relation $\simrel : \vars_\impl \to \vars_{\prog_\classes}$.
\item For each $(\loc,\var) \in \locs \times \vars_\impl$ {\em exactly one} local repair is selected.
\item Each selected local repair $\repair = (\parmap, -) \in \PM$ is \emph{consistent} with $\simrel$, i.e., $\parmap \subseteq \simrel$.
\end{enumerate}
  
The objective function
$\objective = \min \left( \sum_{\repair \in \PM} \mvar{cost}(\repair) \cdot x_\repair \right)$
ensures that the sum of the costs of the selected local repairs is minimal.

Let $\assgn : \ints \to \{ 0, 1 \}$ be a solution of the ILP problem.
We obtain the following total variable relation from $\assgn$: 
$\simrel(\var_2)=\var_1$ iff $\assgn(x_{\var_1\var_2}) = 1$.
For $\assgn(x_\repair) = 1$, where $\PM(\loc, \var) = \repair$, we set $R(\loc,\var) = \repair$.

\paragraph{Adding and Deleting Variables}
The repair algorithm described so far does not change the set of variables,
i.e., $\vars_\repaired = \vars_\impl$.
However, since the repair algorithm constructs a bijective variable relation,
this only works when the implementation and cluster representative have the
same number of variables, i.e., $|\vars_{\impl}| = |\vars_{\prog_\classes}|$.
Hence, we extend the algorithm to also allow the addition and
deletion of variables.

We extend total variable relations $\simrel : \vars_\impl \to \vars_{\prog_\classes}$ to
relations $\simrel \subseteq (\vars_\impl \cup \{ \star \}) \times (\vars_{\prog_\classes} \cup \{ - \})$.
We relax the condition about $\simrel$ being total and bijective:
$\star$ and $-$ can be related to multiple variables or none.
When some variable $\var \in \vars_{\prog_\classes}$ is related to $\star$,
that is $\simrel(\star) = \var$, it denotes that a \emph{fresh variable is added}
to $\pimpl$, in order to match $\var$.
Conversely, when some variable $\var \in \vars_\impl$ is related to $-$, that is $\simrel(\var) = -$,
variable $\var$ is \emph{deleted} from $\pimpl$, together with all its assignments.

Examples of repairs where a fresh variable is added is given in~\figref{fresh-var}
and~\figref{reverse-branches}.

\paragraph{Completeness of the algorithm}
We point out that with this extension the repair algorithm is \emph{complete}
(assuming $\pimpl$ and $\prog_\classes$ have the same control-flow).
This is because the repair algorithm can always generate a \emph{trivial repair}:
all variables $\var_2 \in \vars_\impl$ are deleted, that is
$\simrel(\var_2) = -$ for all $\var_2 \in \vars_\impl$;
and a fresh variable is introduced for every variable $\var_1 \in \vars_{\prog_\classes}$,
that is $\simrel(\star) = \var_1$ for all $\var_1 \in \vars_{\prog_\classes}$.
Clearly, this trivial repair has high cost, and in practice it is very rarely
generated, as witnessed by our experimental evaluation in the next section.

%% file: experiments.tex
\section{Implementation and Experiments}\seclabel{experiments}

\begin{table*}[ht]
  \begin{center}
  \caption{List of the problems with evaluation results for the MOOC data (with AutoGrader comparison).}
  \tablabel{ProblemsAutoGrader}
  \scriptsize
  \begin{tabular}{|c|c|c|c|c|c|c|c|c|c|} \hline
    \textbf{Problem} &
    \textbf{LOC} &
    \textbf{AST size} &
    \multirow{2}{*}{\textbf{\# correct}} &
    \textbf{\# clusters} &
    \multirow{2}{*}{\textbf{\# incorrect}} &
    \multicolumn{2}{c|}{\textbf{\# repaired} (\% of \# incorrect)} &
    \multicolumn{2}{c|}{\textbf{avg. (median) time} in s} \\
    \cline{7-8}
    \cline{9-10}
    \textbf{name} & \textbf{median} & \textbf{median} & &
    (\% of \# correct) & &
     \clara & AutoGrader & \clara & AutoGrader
    \\ \hline
    \noalign{\smallskip} \hline
    \input{table_mitx_all}
  \end{tabular}
  \end{center}
\end{table*}

We now describe our implementation (\secref{implementation})  and
an experimental evaluation, which consists of two parts:
(I)~an evaluation on MOOC data (\secref{MOOCeval}), and
(II)~a user study about the usefulness of the generated repairs (\secref{UserStudy}).
The evaluation was performed on a server with an {\em AMD Opteron 6272 2.1GHz} processor and {\em 224 GB} RAM.

\subsection{Implementation}\seclabel{implementation}
We implemented the proposed approach in the publicly available tool \textsc{Clara}\footnote{\url{https://github.com/iradicek/clara}}
(for {\bf CL}uster {\bf A}nd {\bf R}ep{\bf A}ir).
The tool currently supports programs in the programming languages \clang and \pylang, and consists of:
(1)~Parsers for \clang and \pylang that convert programs to our internal program representation;
(2)~Program and expression evaluation functions for \clang and \pylang,
used in the matching and repair algorithms;
(3)~Matching algorithm;
(4)~Repair algorithm;
(5)~Simple feedback generation system.
We use the {\em lpsolve}~\cite{lpsolve}
{\em ILP} solver, and the {\em zhang-shasha}~\cite{zhang-shasha}
tree-edit-distance algorithm.

\paragraph{Feedback generation}
We have implemented a simple feedback generation system that generates
the location and a textual description of the required modifications
(similar to AutoGrader).
Other types of feedback can be generated from the repair as well, and we
briefly discuss it in~\secref{conclusion}.

\subsection{MOOC Evaluation}\seclabel{MOOCeval}

\paragraph{Setup}
In the first experiment we evaluate \clara on data from the MITx
introductory programming MOOC~\cite{mitxmooc},
which is similar to the data used in evaluation of
AutoGrader~\cite{singh:pldi13}.

This data is stripped from all information about student
identity, i.e., there are not even anonymous identifiers.
To avoid the threat that a student's attempt is repaired by her own future
correct solution, we split the data into two sets.
From the first (chronologically earlier) set we take only the correct
solutions: these solutions are then clustered, and the obtained clusters are
used during the repair of the incorrect attempts.
From the second (chronologically later) set we take only the incorrect attempts: on
these attempts we perform repair.
We have split the data in $80:20$ ratio since then we have a large enough
number ($12973$; see the discussion below) of correct solutions that our
approach requires, while still having quite a large number ($4293$) of
incorrect attempts for the repair evaluation.
We point out that this is precisely the setting that we envision our approach
to be used in: a large number of existing correct solutions (e.g., from a
previous offering of a course) are used to repair new incorrect student
submissions.

\paragraph{Results}
The evaluation summary is in~\tabref{ProblemsAutoGrader};
the descriptions of the problems are in the appendix~\secref{appendix-problems}.
\clara \textbf{automatically} generates a repair for \emph{97.44\%} of
attempts.
As expected, \clara can generate repairs in almost all the cases, since there
is always the \emph{trivial} repair of completely replacing the student
implementation with some correct solution of the same control flow.
Hence, it is mandatory to study the quality and size of the generated repairs.
We evaluate the following questions in more detail:
\emph{(1)~What are the reasons when \clara fails?}
\emph{(2)~Does \clara generate non-trivial repairs?}
\emph{(3)~What is the quality and size of the generated repairs?}

We discuss the results of this evaluation below, while further examples
can be found in the appendix~\secref{more-examples}.

\paragraph{(1) \clara fails}
\clara fails to generate repair in 110 cases:
in 69 cases there are unsupported \pylang features (e.g., nested function definitions),
in 35 cases there is no correct attempt with the same control-flow,
and in 6 cases a numeric precision error occurs in the ILP solver.
The only fundamental problem of our approach is the inability to generate
repairs without matching control-flow; however, since this occurs very rarely,
we leave the extension for a future work.
\emph{Hence, we conclude that CLARA can repair almost all programs.}

\begin{figure}[ht]
  \includegraphics[width=0.7\columnwidth]{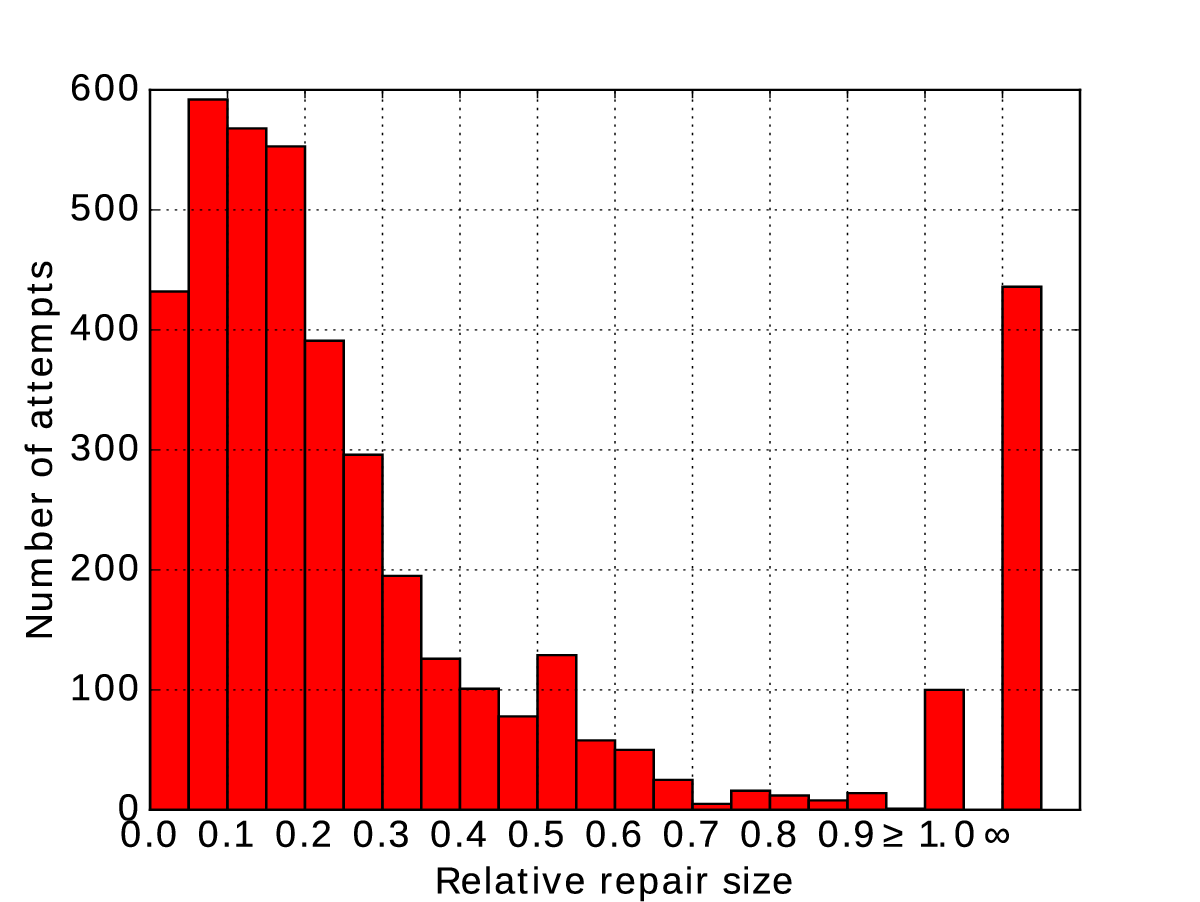}
  \caption{Histogram of relative repair sizes.}
  \figlabel{HistogramRelSize}
\end{figure}

\paragraph{(2) Non-trivial repairs}
Since a correct repair is also a \emph{trivial} one that \emph{completely replaces} a student's
attempt with a correct program, we measure \emph{how much a repair changes
 the student's program}.
To measure this we examine the \emph{relative repair size}:
the \emph{tree-edit-distance} of the repair divided by \emph{the size of the AST} of the program.
Intuitively, the tree-edit-distance tells us how many changes were made in a program,
and normalization with the total number of AST nodes gives us the \emph{ratio of how much of the whole program changed.}
Note, however that this ratio can be $>1.0$, or even $\infty$ if the program is empty.
\figref{HistogramRelSize} shows a histogram of relative repair sizes.
We note that $68\%$ of all repairs have relative size $<0.3$,
$53\%$ have $<0.2$ and $25\%$ have $<0.1$;
the last column ($\infty$) is caused by $436$ completely empty student attempts.
As an example, the two repairs in \figref{MotivExamples}~(g)~and~(h), have
relative sizes of $0.03$ and $0.24$.
\emph{We conclude that \clara in almost all cases generates a non-trivial
  repair that is not a replacement of the whole student's program.}

\begin{figure}[ht]
  \centering
  \scriptsize
  \begin{tabular}{cc}
    \includegraphics[width=0.23\textwidth]{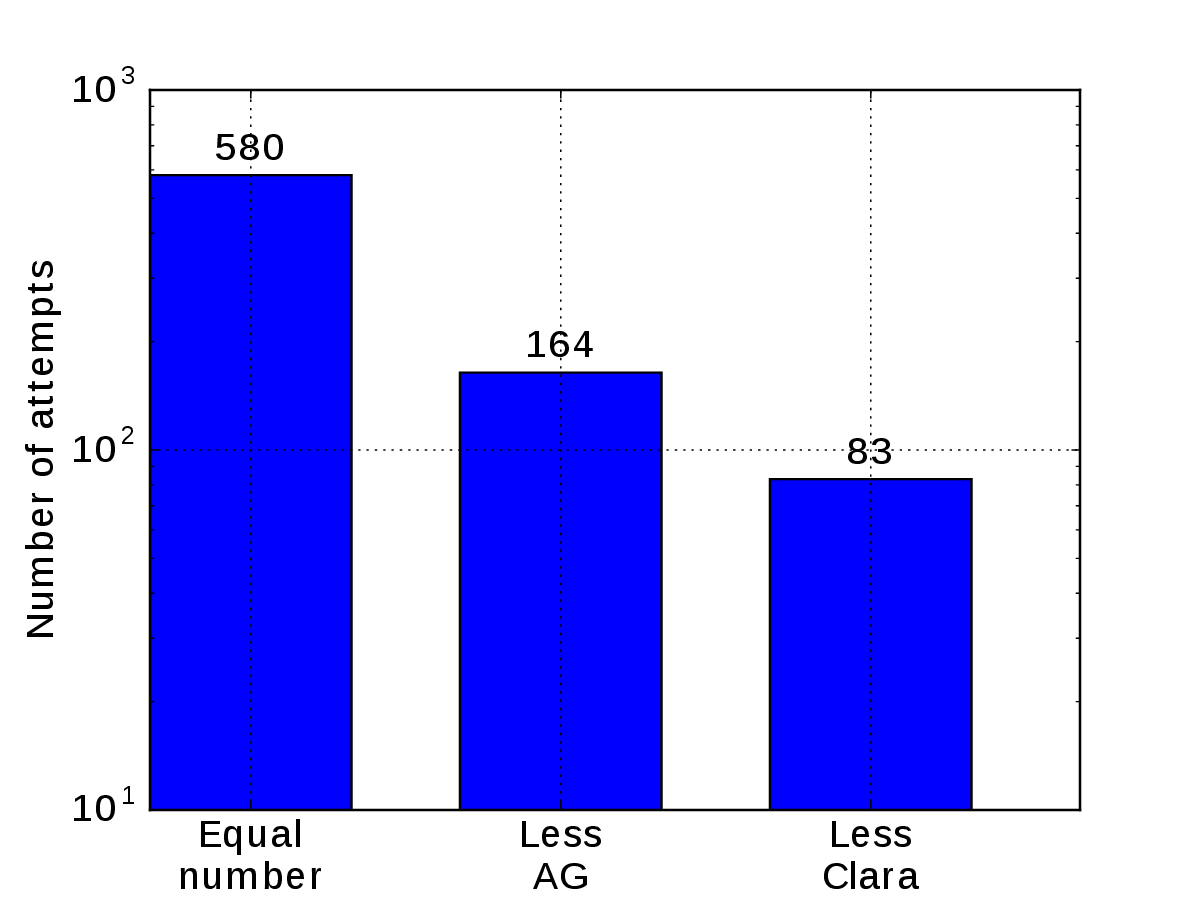} &
    \includegraphics[width=0.23\textwidth]{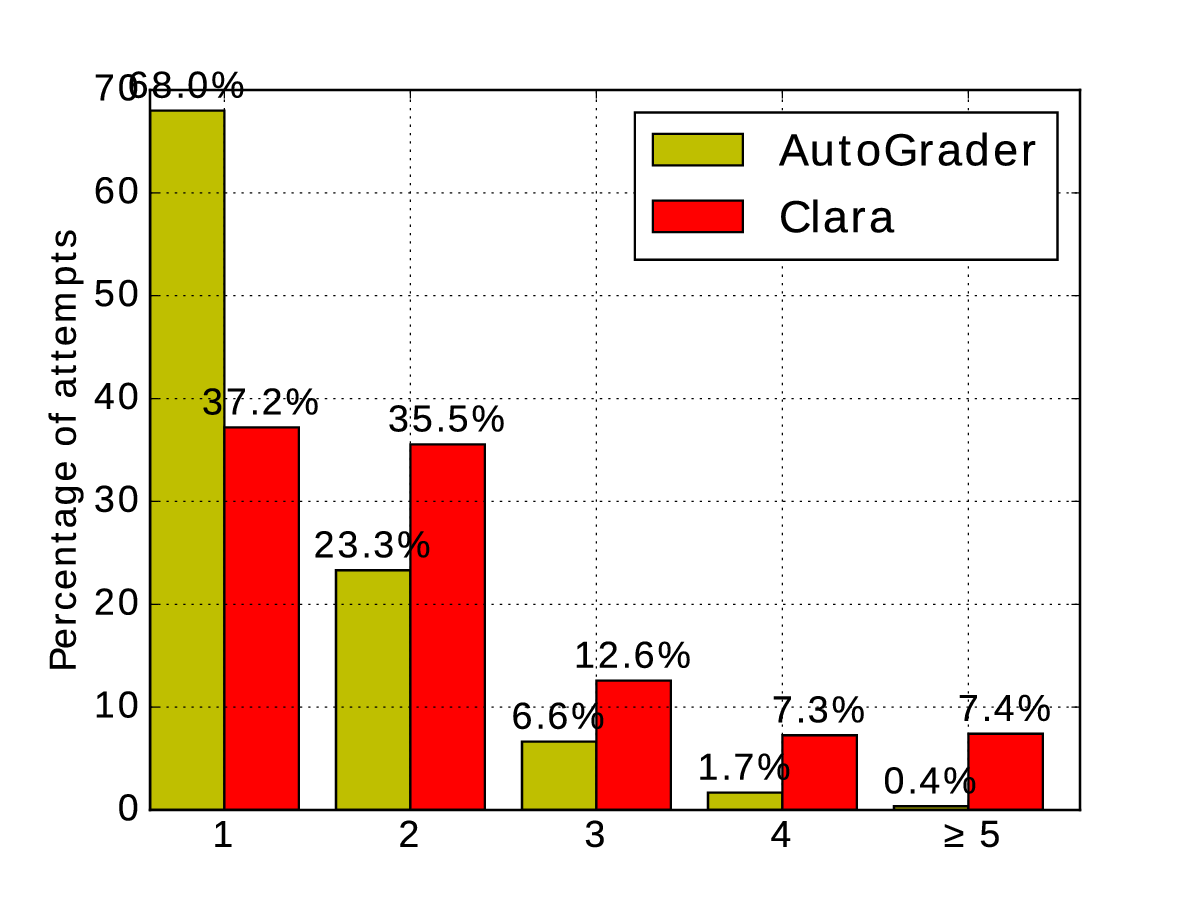} \\
    \begin{minipage}{0.2\textwidth}
      (a) The number of modified expressions per repair.
    \end{minipage} &
    \begin{minipage}{0.2\textwidth}
      (b) Distribution of the number of modified expressions per repair.
    \end{minipage}
  \end{tabular}
  \caption{Comparison of the generated repairs size between AutoGrader and \clara.}
  \figlabel{Plots}
\end{figure}

\paragraph{(3) Repair quality and repair size}
We inspected 100 randomly selected generated repairs, with the goal of
evaluating their quality and size.
Our approach of judging repair quality and size mirrors a human teacher helping a
student: the teacher has to guess the student's idea and use subjective
judgment on what feedback to provide.
We obtained the following results:
(a)~In 72 cases \clara generates \emph{the smallest, most natural repair};
(b)~In 9 cases the repair is almost the smallest, but involves an additional
modification that is not required;
(c)~In 11 cases we determined the repair, although correct, to be different
from the student's idea;
(d)~In 8 cases it is not possible to determine the idea of the student's
attempt, although \clara generates some correct repair.

For the cases in (d), we found that program repair is not adequate and further
research is needed to determine what kind of feedback is suitable when the
student is far from any correct solution.
For the cases in (c), the set of correct solutions does not contain any
solution which is syntactically close to the student's idea; we conjecture that
\clara's results in these cases can partially be improved by considering
different cost functions which do not only take syntactic differences into
account but also make use of semantic information (see the discussion
in~\secref{conclusion}).
However, in 81 cases (the sum of (a) and (b)), \clara \emph{generates good
  quality repairs}.
\emph{We conclude that \clara mostly produces good quality repairs.}

\paragraph{Summary} \emph{Our large-scale experiment on the MOOC data-set shows that \clara can
  fully automatically repair almost all programs and the generated repairs are of
  high quality.}

\paragraph{Clusters}
Finally, we briefly discuss the correct solutions, since our approach depends
on their existence.
The quality of the generated repair should increase with the number of clusters,
since then the algorithm can generate more diverse repairs.
Thus, it is interesting to note that we experienced {\em no performance issues
  with a large number of clusters}; e.g., on \inlinecode{derivatives},
with {\em 532} clusters, a repair is generated on average in {\em 4.9s}.
This is because the repair algorithm processes multiple clusters in parallel.
Nonetheless, clustering is important for \emph{repair quality}, since it
enables repairs that combine expressions taken from different correct solutions
from the same cluster, which would be impossible without clustering.
We found that $2093$ (around $50\%$) repairs were generated using at least
two different correct solutions, and $110$ (around $3\%$) were generated using
at least three different correct solutions, in the same cluster.


\subsubsection{Comparison with AutoGrader}

While the setting of AutoGrader is different (a teacher has to provide an error
model, while our approach is fully automated), the same high-level goal
(finding a minimal repair for a student attempt to provide feedback) warrants
an experimental comparison between the approaches.

\paragraph{Setup and Data}
We were not able to obtain the data used in AutoGrader's evaluation, which
stems from an internal MIT course, because of privacy concerns regarding
student data.
Hence, we compare the tools on the same MITx introductory programming MOOC
data, which we used in the paper for \clara evaluation.
This dataset is similar to the dataset used in AutoGrader's evaluation.
AutoGrader's authors provided us with an AutoGrader version that
is optimized to scale to a MOOC, that is, it has a weaker error
model than in the original AutoGrader's publication~\cite{singh:pldi13}.
According to the authors, some error rewrite rules were intentionally
omitted, since they are too slow for interactive online feedback generation.

\paragraph{Results}
The evaluation summary is in~\tabref{ProblemsAutoGrader}.
AutoGrader is able to generate a repair for \emph{19.29\%} of
attempts, using \emph{manually} specified rewrite-rules,
compared to \emph{97.44\% automatically} generated repairs by \clara.
(We note that AutoGrader is able to repair fewer attempts than reported in the
original publication~\cite{singh:pldi13} due to the differences discussed in
the previous paragraph.)
As \clara can generate repairs in almost all the cases, these numbers are not meaningful on their own; 
the numbers are, however, meaningful in conjunction with our evaluation of the following questions:
\emph{(1) How many repairs can one tool generate that other cannot,
  and what are the reasons when AutoGrader fails?}
\emph{(2) What are the sizes of repairs?}
\emph{(3) What is the quality of the generated repairs, in case both tools generate a repair?}

We summarize the results of this evaluation below, while a more detailed discussion can be found in
the appendix~\secref{more-examples}.

\paragraph{(1) Repair numbers}
In all but one case, when \clara fails to generate a repair, AutoGrader also
fails.
Further, we manually inspected 100 randomly selected cases where AutoGrader
fails, and determined that in 77 cases there is a fundamental problem with
AutoGrader's approach: The modifications require fresh variables, new
statements or larger modifications, which are beyond AutoGrader's capabilities.
\emph{This shows that \clara can generate more complicated repairs than AutoGrader.}

In the 100 cases we manually inspected we also
determined that
\emph{in 74 cases \clara generates good quality repairs, when AutoGrader fails.}

\paragraph{(2) Repair sizes}
We do not report the relative repair size metric for AutoGrader, because we
were not able to extract the repair size from its (textual) output.
However, \figref{Plots}~(a) compares the relation of the
number of modified expressions when both tools generate a repair.
We note that the number of modified expressions is a weaker metric than the
tree-edit-distance, however, we were only able to extract this metric for the
repairs generated by AutoGrader.
\emph{We conclude that AutoGrader produces smaller repair in around 10\% of the
  cases.}

\figref{Plots}~(b) also compares the overall (not just when both tools generate a repair)
distribution of the number of changed expression per repair.
\emph{We notice that most of AutoGrader's repairs modify a single expression,
  and the percentage falls faster than in \clara's case.}

\begin{table*}[ht]
  \begin{center}
  \caption{List of the problems with evaluation details for user study.}\tablabel{ProblemsSurvey}
  \scriptsize
  \begin{tabular}{|c|c|c|c|c|c|c|c|c|c|} \hline
    \multirow{2}{*}{\bf Problem} &
    {\bf LOC} &
    {\bf \# correct} &
    {\bf \# clusters} &
    \multirow{2}{*}{\bf \# incorr.} &
    {\bf \# feedback} &
    {\bf \# repair feedback} &
    \multicolumn{2}{c|}{{\bf time} (in s)} &
    {\bf \# grades} \\
    \cline{8-9}
    & {\bf median}
    & (exist.+study)
    & (exist.+study)
    & & (\% of \# incorr.)
    & (\% of \# feedback)
    & {\bf avg.} & {\bf median}
    & {\bf 1/2/3/4/5 }
    \\ \hline
    \noalign{\smallskip} \hline
    \input{table_survey}
  \end{tabular}
  \end{center}
\end{table*}

\paragraph{(3) Repair quality}
Finally, we manually inspected 100 randomly selected cases where
both tools generate a repair.
In 61 cases we found both tools to produce the same repair;
in 19 cases different, although of the same quality;
in 9 cases we consider AutoGrader to be better;
in 5 cases we consider \clara~to be better;
and in 6 cases we found that AutoGrader generates an incorrect
repair.
\emph{We conclude that there is no notable difference between the tools when
  both tools generate a repair.}


\subsection{User Study on Usefulness}\seclabel{UserStudy}

In the second experiment we performed a user study, evaluating \clara
in real time.  We were interested in the following questions:
(1) {\em How often and fast is feedback generated (performance)?}
(2) {\em How useful is the generated repair-based feedback?}

\paragraph{Setup}
To answer these questions we developed a web interface for \clara and conducted
a user study, which we advertised on programming forums, mailing lists, and
social networks.
Each participant was asked to solve six introductory \clang programming problems, for
which the participants received feedback generated by \clara.
There was one additional problem, not discussed here, that was almost solved,
and whose purpose was to familiarize the participants with the interface.
After solving a problem each participant was
presented with the question: {\em ``How useful was the feedback provided on this
  problem?''}, and could select a grade on the scale from 1 ("{\em Not useful
  at all}") to 5 ("{\em Very useful}").
Additionally, each participant could enter an additional textual comment for
each generated feedback individually and at the end of solving a problem.

We also asked the participants to assess their programming experience
with the question: {\em ``Your
  overall programming experience (your own, subjective, assessment)''}, with
choices on the scale from 1 (``{\em Beginner}'') to 5 (``{\em Expert}'').

The initial correct attempts were taken from an introductory programming
course at {\em IIT Kanpur, India}.
The course is taken by around 400 students of whom several
have never written a program before.
We selected problems from two weeks where students start solving more
complicated problems using loops.
Of the 16 problems assigned in these two weeks, we picked those 6 that were
sufficiently different.

\paragraph{Results}
\tabref{ProblemsSurvey} shows the summary of the results;
detailed descriptions of all problems are available in the appendix~\secref{appendix-problems}.
The columns {\bf \# correct} and {\bf \# clusters} show
the number of correct attempts and clusters obtained from:
(a) the existing ESC 101 data ({\em exist.} in the table), and
(b) during the case study from participants' correct attempts ({\em study} in the table).
We plan to make the complete data, with all attempts, grades and textual
comments publicly available.

\paragraph{Performance of \clara}
Feedback was generated for 1503 ({\bf 88.52\%}) of incorrect attempts.
In the following we discuss the 3 reasons why feedback could not be generated:
(1)~In 57 cases there was a bug in \textsc{Clara}, which we have fixed after
the experiment finished.
Then we confirmed that in all 57 cases the program is correctly repaired and
feedback is generated (note that this bug was only present in this real-time
experiment, i.e., it did not impact the experiment described in the previous
section);
(2)~In 43 cases a timeout occurred (set to 60s);
(3)~In 95 cases a program contained an unsupported C construct, or there was a
syntactic compilation error that was not handled by the web interface (\clara
currently provides no feedback on programs that cannot be even parsed).
Further, the average time to generate feedback was \textbf{8} seconds.
\emph{These results show that \textsc{Clara} provides feedback on a large
  percentage of attempts in real time.}

\paragraph{Feedback usefulness}
The results are based on {\em 191} grades given by {\em 52} participants.
Note that
problems have a different number of grades.
This is because we asked for a grade only when feedback was successfully
generated (as noted above, in $88.52\%$ cases), and because some of the
participants did not complete the study.
The average grade over all problems is {\bf 3.4}.
{\em This shows very promising preliminary results on the usefulness of \textsc{Clara}}.
However, we believe that these results can be further improved (see \secref{conclusion}).

The participants declared their experience as follows: 22 as 5, 19 as 4, 9 as
3, 0 as 2, and 2 as 1.  While these are useful preliminary results, a study
with beginner programmers is an important future work.

\emph{Note.} In the case of a very large repair (\mbox{$\mathit{cost}>100$} in our study),
we decided to show a generic feedback explaining a general strategy
on how to solve the problem.
This is because the feedback generated by such a large repair is usually
not useful.
We generated such a general strategy in 403 cases.

\subsection{Threats to Validity}

\paragraph{Program size}
We have evaluated our approach on small to medium size programs typically found
in introductory programming problems.
The extension of our approach to larger programming problems, as found in more
advanced courses, is left for future work.
Focusing on small to medium size programs is in line with related work on
automated feedback generation for introductory programming (e.g.,
\citet{qlose}, \citet{singh:pldi13}, \citet{LS17}).
We stress that the state-of-the-art in teaching is manual feedback (as well as
failing test cases); thus, automation, even for small to medium size programs,
promises huge benefits.
We also mention that our dataset contains larger and challenging attempts by
students which use \emph{multiple functions}, \emph{multiple} and
\emph{nested loops}, and our approach is able to handle them.

The focus of our work differs from the related work on program repair
(see~\secref{related} for a more detailed discussion).
Our approach is specifically designed to perform well on small to medium size
programs typically found in introductory programming problems, rather than the
larger programs targeted in the literature on automated program repair.
In particular, our approach addresses the challenges of (1) a high number of
errors (education programs are expected to have higher error
density~\cite{singh:pldi13}), (2) complex repairs, and (3) a runtime fast enough
for use in an interactive teaching environment.
These goals are often out of reach for program repair techniques.
For example, the general purpose program repair techniques discussed
in~\citet{IntroClass} on the IntroClass benchmark, either repair a small number
of defects (usually <50\%) or take a long time (i.e., over one minute).

\paragraph{Unsoundness}
Our approach guarantees only that repairs are correct over a given set of test
cases.
This is in accordance with the state-of-the-art in teaching, where testing is
routinely used by course instructors to grade programming assignments and
provide feedback (e.g., for ESC101 at IIT Kanpur, India~\cite{prutor}).
When we manually inspected the repairs for their correctness, we did not find
any problems with soundness.
We believe that this due to the fact that programming problems are small,
human-designed problems that have comprehensive sets of test cases.

In contrast to our dynamic approach, one might think about a sound static approach based on
symbolic execution and SMT solving.
We decided for a dynamic analysis because symbolic execution can sometimes take
a long time or even fail when constructs are not supported by an SMT solver.
For example, reasoning about floating points and lists is difficult for SMT
solvers.
On the other hand, our method only executes given expressions on a set of
inputs, so we can handle any expression, and our method is fast.
Further, our evaluation showed our dynamic approach to be precise enough for the
domain of introductory programming assignments.
The investigation of a static verification of the results generated by our
repair approach is an interesting direction for future work:
one could take the generated repair expressions and verify that they indeed
establish a simulation with the cluster against which the program was repaired.

%% file: table_mitx_all.tex
\inlinecode{derivatives} & $14$ & $33$ & $1472$ & $532$ ($36.14\%$) & $481$ & $472$ ($98.13\%$) & $235$ ($48.86\%$) & $4.9$s ($4.4$s) & $6.6$s ($5.2$s) \\ 
\inlinecode{oddTuples} & $10$ & $25$ & $9001$ & $454$ ($5.04\%$) & $3584$ &  $3514$ ($98.05\%$) & $576$ ($16.07\%$) & $3.0$s ($2.6$s) & $25.5$s ($13.3$s) \\ 
\inlinecode{polynomials} & $13$ & $25$ & $2500$ & $234$ ($9.36\%$) & $228$ & $197$ ($86.40\%$) & $17$ ($7.46\%$) & $1.9$s ($1.6$s) & $4.3$s ($4.0$s) \\ 
\hline Total &$11$ & $25$ & $12973$ & $1220$ ($9.40\%$) & $4293$ & $4183$ ($97.44\%$) & $828$ ($19.29\%$) & $3.2$s ($2.7$s) & $19.7$s ($6.3$s) \\ 
\hline

%% file: table_survey.tex
\inlinecode{Fibonacci sequence} & 12 & 512+84 & 70 + 17 (14.60\%) & 572 & 539 (94.23\%) & 440 (81.63\%) & 10.44 & 8.51 &  1 / 7 / 9 / 16 / 13 \\ \hline
\inlinecode{Special number} & 15 & 358+59 & 39 + 3 (10.07\%) & 121 & 109 (90.08\%) & 94 (86.24\%) & 3.77 & 2.38 &  2 / 3 / 8 / 9 / 13 \\ \hline
\inlinecode{Reverse Difference} & 17 & 342+46 & 48 + 8 (14.43\%) & 103 & 77 (74.76\%) & 68 (88.31\%) & 4.39 & 3.07 &   4 / 4 / 5 / 3 / 5 \\ \hline
\inlinecode{Factorial interval} & 14 & 391+44 & 56 + 8 (14.71\%) & 234 & 232 (99.15\%) & 185 (79.74\%) & 3.33 & 3.17 &  2 / 5 / 4 / 5 / 13 \\ \hline
\inlinecode{Trapezoid} & 14 & 281+41 & 36 + 15 (15.84\%) & 143 & 129 (90.21\%) & 121 (93.80\%) & 7.55 & 4.82 &   7 / 5 / 7 / 7 / 5 \\ \hline
\inlinecode{Rhombus} & 21 & 264+38 & 73 + 22 (31.46\%) & 525 & 417 (79.43\%) & 192 (46.04\%) & 9.16 & 5.35 &  6 / 9 / 6 / 5 / 3 \\ \hline

%% file: related.tex
\section{Related Work}\seclabel{related}

\paragraph{Automated Feedback Generation}
\citet{Ihantola:2010} \\
present a survey of
tools for the automatic assessment of programming exercises.
Pex4Fun~\cite{tillmann13:teaching} and its successor CodeHunt~\cite{tillmann:2014}
are browser-based, interactive platforms
where students solve programming assignments with hidden specifications, and
are presented with a list of automatically generated test cases.
LAURA~\cite{Adam:1980} heuristically applies program transformations to a student's
program and compares it to a reference solution, while reporting mismatches
as potential errors (they could also be correct variations).
Apex~\cite{apex} is a system that automatically generates error explanations
for bugs in assignments, while our work automatically clusters
solutions and generates repairs for incorrect attempts.

\paragraph{Trace analysis}
\citet{striewe:11}~have proposed presenting full program traces to the students,
but the interpretation of the traces is left to the students.
They have also suggested automatically comparing the student's trace to that of a sample
solution~\cite{striewe:13}.
However, the approach misses a discussion of the situation when the student's code enters an infinite loop,
or has an error early in the program that influences the rest of the trace. 
The approach of~\citet{our:fse14} uses a dynamic analysis based approach to find
a strategy used by the student, and to generate feedback for {\em performance aspects}.
However, the approach requires specifications {\em manually provided} by the teacher,
written in a specially designed specification language, and it
only matches specifications to correct attempts, i.e.,
it {\em cannot provide feedback on incorrect attempts}.

\paragraph{Program Classification in Education}
CodeWebs~\cite{Nguyen:2014} classifies different AST sub-trees in equivalence classes
based on probabilistic reasoning and program execution on a set of inputs.
The classification is used to build a search engine over ASTs to enable
the instructor to search for similar attempts, and to provide feedback on some class of ASTs.
OverCode~\cite{Glassman:2014} is a visualization system that uses
a lightweight static and dynamic analysis, together with manually provided rewrite rules, to group student attempts.
\citet{Drummond:2014}~propose a statistical approach to classify
interactive programs in two categories ({\em good} and {\em bad}).
\citet{LS17} cluster incorrect programs by the type of the required
modifications.
CoderAssist~\cite{SemiSupervisedFSE16} provides feedback on student
implementations of dynamic programming algorithms: the approach first clusters
both correct and incorrect programs based on their syntactic features; feedback
for incorrect program is generated from a counterexample obtained from
an equivalence check (using SMT) against a correct solution in the same cluster.

\paragraph{Program Repair}
The research on program repair is vast; we mention some work, with emphasis on
introductory education.
The non-education program repair approaches are based on SAT~\cite{Gopinath:2011},
symbolic execution~\cite{Konighofer:2011},
games~\cite{Jobstmann:2005,Staber:2005},
program mutation~\cite{Debroy:2010},
and genetic programming~\cite{Arcuri:2008, Forrest:2009}.
In contrast, our approach uses dynamic analysis for scalability.
These approaches aim at repairing large programs, and therefore are not able to
generate complex repairs.
Our approach repairs small programs in education and uses multiple correct
solutions to find the best repair suggestions, and therefore is able to suggest
more complex repairs.

Prophet~\cite{Prophet} mines a database of successful patches and uses
these patches to repair defects in large, real-world applications.
However, it is unclear how this approach would be applicable to our educational setting.
SearchRepair~\cite{SearchRepair} mines a body of code for short snippets that it
uses for repair.
However, SearchRepair has different goals than our work and has not been used
or evaluated in introductory education.
Angelic Debugging~\cite{AngelicDebugging} is an approach that
identifies \emph{at most one} faulty expression in the program and tries to
replace it with a correct \emph{value} (instead of \emph{replacement expression}).

\citet{FeasibilityFSE16} explore different automated program repair (APR)
systems in the context of generating feedback in intelligent tutoring systems.
They show that using APR out-of-the-box seems infeasible due to the low
repair rate, but discuss how these systems can be used to generate
\emph{partial repairs}.
In contrast, our approach is designed to provide complete repairs.
They also conclude that further research is required to understand how to
generate the most effective feedback for students from these repairs.

AutoGrader~\cite{singh:pldi13} takes as input an incorrect student program,
along with a reference solution and a set of potential corrections in the form
of expression rewrite rules (both provided by the course instructor), and
searches for a set of minimal corrections using program synthesis.
In contrast, our approach is completely automatic and can generate more
complicated repairs.
\textsc{Refazer}~\cite{Refazer} learns programs transformations from example
code edits made by the students, and then uses these transformations to repair
incorrect student submissions.
In comparison to our approach, \textsc{Refazer} does not have a cost model, and
hence the generated repair is the first one found (instead of the smallest
one).
\citet{Rivers17} transform programs to a canonical form using
semantic-preserving syntax transformations, and then report syntax difference
between an incorrect program and the closest correct solution; the paper
reports evaluation on loop-less programs.
In contrast, our approach uses \emph{dynamic equivalence}, instead of
(canonical) syntax equivalence, for better robustness under syntactic
variations of semantically equivalent code.
\textsc{Qlose}~\cite{qlose} automatically repairs programs in education
based on different program distances.
The idea to consider different semantic distances is very interesting,
however the paper reports only a very small initial evaluation (on 11 programs),
and \textsc{Qlose} is only able to generate small, template-based repairs.


%% file: futurework.tex
\section{Future Work}\seclabel{futurework}

In this section we briefly discuss the limitations of our approach,
and possible directions for future work.

\paragraph{Cost function}
The cost function in our approach compares only the syntactic difference
between the original and the replacement expressions (specifically, we use the
tree-edit-distance in our implementation).
We believe that the cost function could take into account more information;
e.g., variable roles~\cite{variableroles} or semantic distance~\cite{qlose}.

\paragraph{Control-flow}
The clustering and repair algorithms are restricted to the analysis of programs
with the same control-flow.
As the case of ``no matching control-flow to generate a repair'' rarely occurs in
our experiments (only 35 cases in the MOOC experiment), we have left the
extension of our algorithm to programs with different control-flow for future
work.
We conjecture that our algorithm could be extended to programs with similar
control-flow (e.g., different looping-structure).

\paragraph{Feedback}
Our tool currently outputs a textual description of the generated repair, very
similar to the feedback generated by AutoGrader.
We believe that the generated repairs could be used to derive other types of feedback as well.
For example, a more abstract feedback with the help of a course instructor:
A course instructor could annotate variables in the correct solutions with
their descriptions, and when a repair for some variable is required, a matching
feedback is shown to a student.

While this paper is focused on the technical problem of finding possible
repairs, an interesting orthogonal direction for future work is to consider
pedagogical research questions, for example:
(1)~How much information should be revealed to the student (the line number, an incorrect expression, the whole repair)?
(2)~Should the use of automated help be penalized?
(3)~How much do students learn from the automated help?

%% file: conclusion.tex
\section{Conclusion}\seclabel{conclusion}

We present novel algorithms for clustering and program repair in introductory
programming education.
The key idea behind our approach is to use the \emph{existing correct student
solutions}, which are available in tens of thousands in large MOOC courses, to
\emph{repair incorrect student attempts}.
Our evaluation shows that \clara can generate a large number of repairs without
any manual intervention, can perform complicated repairs, can be used in an
interactive teaching setting, and generates good quality repairs in a large percentage of cases.

%% file: appendix-problems.tex
\section{List of Problems in the Evaluation}\seclabel{appendix-problems}

Here we list the descriptions of the problems used in our evaluation.

\paragraph{MOOC evaluation:}

\begin{itemize}
\item {\inlinecode{derivatives}}
  
  Compute and return the derivative of a polynomial function as a list of floats.
  If the derivative is $0$, return $[0.0]$.

  input: list of numbers (length $\geq 0$)
  
  return: list of numbers (floats)
  \\

\item \inlinecode{oddTuples}

  input: a tuple aTup

  return: a tuple, every other element of aTup.
  \\

\item \inlinecode{polynomials}
  
  Compute the value of a polynomial function at a given value $x$. Return that value as a float.

  inputs: list of numbers (length $> 0$) and a number (float)

  return: float

\end{itemize}

\paragraph{User study:}

\begin{itemize}

\item \inlinecode{Fibonacci sequence}

  Write a program that takes as input an integer $k > 0$ and prints the integer $n > 0$ such that $F_n \leq k < F_{n+1}$. \\
  Here $F_n$ means the $n^{\text{th}}$ number in the Fibonacci sequence defined by the relation: \\
  $F_n = F_{n-1} + F_{n-2}$, for $n > 2$ \\
  $F_1 = 1$ \\
  $F_2 = 1$ \\
  Examples of Fibonacci numbers are: $1, 1, 2, 3, 5, 8, \dots$
  \\

\item \inlinecode{Special number}

  Write a program that takes as input an integer $n \ge 0$ and prints $\text{YES}$ if $n$ is a {\em special} number, and $\text{NO}$ otherwise. \\
  A number is special if the sum of cubes of its digits is equal to the number itself. \\
  {\em Note}: A cube of some number $x$ is $x^3 = x \cdot x \cdot x$. \\
  {\em For example}: $371$ is a special number, since $3^3 + 7^3 + 1^3 = 27 + 343 + 1 = 371$.
  \\

\item \inlinecode{Reverse Difference}

  Write a program that takes as input a positive integer $n>0$ and prints the difference of $n$ and its reverse.
  For example, if $n$ is $1234$, the output will be $-3087$ (result of $1234 - 4321$).
  \\

\item \inlinecode{Factorial interval}

  Write a program that takes as input two integers $n$ and $m$ (where $0 \leq n \leq m$), and prints the number ({\em count}) of factorial numbers in the closed interval $[n, m]$. \\
  A number $f$ is a factorial number if there exists some integer $i \ge 0$ such that $f = i!$ \\
  {\em Note}: $i! = 1 \cdot 2 \cdots i$, that is, $i!$ is a product of first $i$ natural numbers, excluding $0$. \\
  Examples of factorial numbers are: $1, 2, 6, 24, 120, \dots$
  \\

\item \inlinecode{Trapezoid}

  Write a program to do the following: \\
  (a) Read height $h$ and base length $b$ as the input; \\
  (b) Print $h$ lines of output such that they form a pattern in the shape of a {\em regular trapezoid}. \\
  (c) Trapezoid should be formed using the symbol "*". \\
  Example output for $h=5$ and $b=14$ ("-" denotes where spaces should go, you should print a space " " instead of "-"): \\
  
\begin{verbatim}
----******
---********
--**********
-************
**************
\end{verbatim}

{\em Important}: There should be \emph{NO ANY EXTRA SPACE} (before the pattern, between rows, between columns,~\dots). The last line should be an empty line.
\\

\item \inlinecode{Rhombus}

  Write a program to do the following: \\
  (a) Take height $h$ as the input; \\
  (b) Print $h$ lines of output such that they form a pattern in the shape of a {\em rhombus}; \\
  (c) Each line should be formed by the integer representing the column number modulo 10. \\
  {\em Note}: You can assume that $h$ will be odd and $h \ge 3$. \\
  Example output for $h=5$ ("-" denotes where spaces should go,  you should print a space " " instead of "-"): \\
  
\begin{verbatim}
--3
-234
12345
-234
--3
\end{verbatim}

\emph{Important}: There should be \emph{NO ANY EXTRA SPACE} (before the pattern, between rows, between columns,~\dots). The last line should be an empty line.
  
\end{itemize}

%% file: appendix-code-examples.tex
\clearpage

\section{Additional Code Examples}\seclabel{more-examples}

Next, we discuss some further examples from the MOOC evaluation and comparison
with AutoGrader.
Specifically, we give examples when
AutoGrader fails, but \clara is able to generate a repair, and
when \clara generates a repair that is almost the smallest repair, but involves an
additional, unnecessary modification.

\begin{figure}[h!]
  \includecode{examples/clara-ok-1.py}{1.0}

  The generated repair is:
  \begin{enumerate}
  \item Add a new variable with assignment \inlinecode{new\_x = 1} at the beginning of function \inlinecode{oddTuples}.
  \item In codition at line 4 change \inlinecode{i.length()\%2 != 0} to \inlinecode{new\_x \%2 != 0}.
  \item Add assignment \inlinecode{new\_x = new\_x + 1} inside the loop starting at line 3.
  \item Add return statement \inlinecode{return tuple} after the loop staring at line 3.
  \end{enumerate}
  The relative cost of this repair is $0.28$. \\
  
  \caption{Big conceptual error.}
  \figlabel{fresh-var}
\end{figure}

AutoGrader cannot generate the repair in \figref{fresh-var} for several reasons:
\begin{itemize}
\item It requires adding a fresh variable, which their error model does not support;
\item It requires adding two new statements (assignment to \inlinecode{new\_x} and a return statement),
which their error model also does not support;
\item It requires changing a whole sub-expression \inlinecode{i.length()} with a variable.
\end{itemize}
The modification (2.) is not possible in AutoGrader (even if we ignore the freshly added variable)
because it requires changing an arbitrary expression with some variable.
AutoGrader's authors describe this as a \emph{big conceptual error} in their paper,
and also mention that this is one of the biggest challenges for AutoGrader.

The next two examples show cases where \clara generates repair that is
correct, but also involves an additional (unnecessary) modification.

\begin{figure}[h!]
  \includecode{examples/clara-reverse-branches.py}{1.0}

  The generated repair is:
  \begin{enumerate}
  \item Add a new variable with assignment \inlinecode{new\_ans=()} to the beginning of function \inlinecode{oddTuples}.
  \item In condition at line 2 change \inlinecode{len(aTup)==1} to \inlinecode{len(aTup)==0}.
  \item In condition at line 4 change \inlinecode{len(aTup)==0} to \inlinecode{len(aTup)==1}.
  \item In the iterator at line 7 change \inlinecode{range(1, len(aTup))} to \inlinecode{range(0, len(aTup), 2)}.
  \item Add assignment \inlinecode{new\_ans = new\_ans + aTup[n]} inside the loop starting at line 7.
  \item Add return statement \inlinecode{return new\_ans} after the loop staring at line 7.
  \end{enumerate}
  The relative cost of this repair is $0.48$. \\
  \caption{Reverse condition branches in the repair.}
  \figlabel{reverse-branches}
\end{figure}

The repair in~\figref{reverse-branches} would also be correct if modifications (2.) and (3.) were omitted, but \clara
generates this repair since the closest correct attempt has the branches reversed,
i.e., it first examines the case when \inlinecode{len(aTup)==0}, and since the repair
algorithm requires the same control-flow, it also suggests these modifications.
To eliminate these two modifications, we would have to relax our repair algorithm's requirement
on control-flow; however it is not clear at the moment how to do that.

On the other hand AutoGrader cannot repair this attempt for two reasons:
\begin{itemize}
  \item It requires addition of a new variable, which is not expressible
    in its error model;
  \item It requires adding 3 new statements (two assignments to a fresh variable \inlinecode{new\_ans},
    and a return statement), which is also not expressible in its error model.
\end{itemize}

\begin{figure}[h!]
  \includecode{examples/clara-nok-2.py}{1.0}

  The generated repair is:
  \begin{enumerate}
  \item Change assignment \inlinecode{rTup = ''} to \inlinecode{rTup = ()} at line 2.
  \item Add assignment \inlinecode{index = 0} in the beginning of function \inlinecode{oddTuples}.
  \item In the iterator at line 3, change \inlinecode{range(0, len(aTup))} to \inlinecode{range(0, len(aTup), 2)}.
  \item Change assignment \inlinecode{rTup += aTup(index)} to \inlinecode{rTup += (aTup[index],)} at line 4.
  \end{enumerate}
  The relative cost of this repair is $0.19$. \\

  \caption{Additional statement.}
  \figlabel{additional-statement}
\end{figure}

The repair in~\figref{additional-statement} would be also correct if the modification (2.) is omitted,
but, same as in the previous example, because it is present in the correct
solution \clara generates this modification as well.
This could be handled by performing an additional analysis that
would find this statement redundant.

However, AutoGrader did not manage to generate any repair for this example.